\documentclass[preprint,amsmath]{revtex4-1}

\usepackage{amsmath}
\usepackage{amssymb}
\usepackage{color}
\usepackage[normalem]{ulem}
\usepackage[justification=centerlast,font={small,singlespacing}]{caption}
\usepackage[justification=centerlast,font={small,singlespacing}]{subcaption}
\usepackage{mathtools}

\begin{document}

\title{Stochastic Maps, Continuous Approximation and Stable Distribution }
\author{David A. Kessler}
\email{kessler@dave.ph.biu.ac.il}
\author{Stanislav Burov}
\email{stasbur@gmail.com}
\affiliation{Physics Department, Bar-Ilan University, Ramat Gan 52900,
Israel}

\pacs{PACS}

\begin{abstract}
A continuous approximation framework for non-linear stochastic as well as deterministic discrete maps is developed. For the stochastic map with uncorelated Gaussian noise, by successively applying the It\^{o} lemma, we obtain a Langevin type of equation. Specifically, we show how non-linear  maps  give rise to a Langevin description that involves multiplicative noise. The multiplicative nature of the noise induces an additional effective force, not present in the absence of noise. We further exploit the continuum description and  provide  an explicit formula for the stable distribution of the stochastic map and conditions for its existence. Our results are in good agreement with numerical simulations of several maps. 
\end{abstract}

\maketitle

\section{Introduction}

 A dynamical description of most physical situations is achieved by means of differential equations. For systems where noise is present, a common dynamical description is by means of a stochastic differential equation (SDE). A well-known example is the drift-diffusion equation describing the dynamics of some variable  $X$ as a function of time $t$, 
 \begin{equation}
 dX_t = \mu_t\,dt+\sigma_t dB_t.
 \label{ito_eq01}
 \end{equation}
Here,  $B_t$ is a Wiener process~\cite{Gardiner}, or, more simply, the noise term (as it commonly called in physics), $\sigma_t$ is the noise variance, and $\mu_t$ is the drift term.  The behavior of SDEs has been extensively explored due to its countless applications in physics, chemistry and economics and many other fields~\cite{Gardiner}. For example, the physical analog of Eq.~(\ref{ito_eq01}), the overdamped Langevin equation, is a common tool for theoretical and computational studies of Brownian motion and thermal noise in electrical resistors~\cite{GitBook}.
Since $X$ is a random variable that fluctuates as a function of time, its values attain a time-dependent distribution ${\cal{P}} _{t}(X)$. The Fokker-Planck 
equation dictates the evolution with time of ${\cal{P}} _{t}(X)$ and for specific types of processes a stable distribution exists, so that
${\cal P}_t(X) \to {\cal P}_\infty(X)$ as $t \to \infty$~\cite{Risken}. The way to find the stable distribution is well established and takes the form of the Maxwell-Boltzmann distribution ${\cal P}_\infty (X) \propto \exp\left( -U(X)/k_B T\right)$. The potential $U(X)$ is defined by the drift term (when it is independent of $t$) and $k_B T$ (Boltzmann constant $\times$ temperature) is proportional to $\sigma_t$ (when the variance is constant).

The discrete-time analogs of differential equations are maps. Maps have been also a subject of extensive research in fields like non-linear dynamics and chaos theory~\cite{Strogatz,Bak}.  Simply speaking, $X_t\xrightarrow{G} X_{t+1}$ is a one-dimensional discrete map such that $X_{t+1}=G(X_t)$, and the functional properties of $G$ determine objects such as fixed points and limit cycles. The discrete analog of an SDE is a stochastic map (SM) 
\begin{equation}
X_{t+1} = G(X_t)+\eta_t,
\label{noisymap01}
\end{equation}
where $\eta_t$ is a general random variable. SMs appeared in the mathematical literature~\cite{Kesten1,Kesten2} quite a while ago and are widely used in financial mathematics as well~\cite{finance1}. The effects of noise on chaotic systems drew attention to SMs as well~\cite{McKane1,McKane2,McKane3,Reimann1,Reimann2,Fogedby}. Recent biological applications, concerned with bacteria growth and protein expression use SMs as basic models~\cite{Amir1,Amir2,Naama1,Tanuchi}. 
Generally, for every process that depends on a discrete parameter and for which intrinsic (or extrinsic) noise is unavoidable, the proper description will be in terms of an SM. 
SMs are in fact relevant  even in the context of continuous times processes described by SDEs. Any numerical method for an SDE is based on simulations of the discretized form of  the SDE~\cite{simLangevin}, a form which usually looks like Eq.~(\ref{noisymap01}) with $G(X_t)=X_t+a(X_t)$, where $a(X_t)$  is some general function. 

While, as mentioned, stochastic maps appear in a quite large variety of disciplines, their treatment is usually restricted  to the known exact solution of a linear map~\cite{McKane2,Amir1,Naama1}. A general approach to stochastic maps is missing, especially in the context of properties of their stable distributions ${\cal P}_\infty(X)$. The goal of this manuscript is to develop a continuous approximation for SMs, an approximation that will result in an SDE  similar to Eq.~({\ref{ito_eq01}). With such an approximation in hand, it will be possible to exploit the well-developed theory of Langevin and Fokker-Planck equations. Specifically, the stability properties of the SM can be deduced from the existence of a stable ${\cal P}_\infty(X)$ as a stationary solution of the Fokker-Planck equation.

We restrict our treatment of SM to the specific case of Gaussian and uncorrelated $\eta_t$ and develop a continuous approximation when $\eta_t$ is independent of $X$. The obtained approximation is applied to several (linear and non-linear) SMs, some of which intentionally don't fit the original restrictions of the approximation method with respect to noise properties. The results show that careful application of the derived approximation leads to a quite satisfactory description of SM stable behavior in terms of the associated continuous SDE.  

\section{Non-Stochastic Map Approximation}
\label{nonstoch}

We start with an approximation method for a noise-free discrete analog of Eq.~(\ref{noisymap01}), 
\begin{equation}
X_{t+1}=X_t+a(X_t). 
\label{secnon01}
\end{equation}
Our continuum approximation assumes that $t$ can take any value in $R$ and  that $X_t(t)$ is differentiable with respect to $t$. The general solution that we seek is of the form 
\begin{equation}
X_t=X_{t_0}+\int_{t_0}^{t}b(X_{t'})\,dt',
\label{secnon02}
\end{equation}
where $b(X_t)$ is some yet unknown function. It is clear that $b(X_t)$ must satisfy, $\forall t$,  
$a(X_t)=\int_t^{t+1}b(X_{t'})\,dt'$ in order for Eq.~(\ref{secnon01}) to be consistent with Eq.~(\ref{secnon02}).
The function $a(X)$ in Eq.~(\ref{secnon01}) is assumed to be infinitely differentiable. Applying the Euler-Maclaurin formula~\cite{Abramowitz},
we obtain for $a(X_t)$,
\begin{equation}
a(X_t) = \int_t^{t+1} a(X_{t'})\,dt' -\frac{1}{2}\int_t^{t+1}\left(\frac{da(X_{t'})}{dt'}\right)\,dt'
+\sum_{k=1}^{\infty} \frac{B_{2k}}{(2k)!}\int_t^{t+1}\frac{d^{2k}a(X_{t'})}{dt'^{2k}}\,dt',
\label{secnon03}
\end{equation}
where $B_i$ are Bernoulli numbers. From Equations (\ref{secnon01}), (\ref{secnon02}) and (\ref{secnon03}) we obtain 
\begin{equation}
b(X_t) = a(X_{t}) -\frac{1}{2}\frac{da(X_{t})}{dt}
+\sum_{k=1}^{\infty} \frac{B_{2k}}{(2k)!}\frac{d^{2k}a(X_{t})}{dt^{2k}}.
\label{secnon04}
\end{equation}
Equation~(\ref{secnon04}) is an expansion for $b(X_t)$. The first order approximation will simply be $b(X_t)=a(X_t)$. 
At second order we get $b(X_t)=a(X_t)-\left(1/2\right)\,da(X_t)/dt$. Using the chain rule for the $t$ derivative of $a(X_t)$ and the 
differential form of Eq.~(\ref{secnon02}), i.e., $d\,X_t/dt = b(X_t)$, we obtain 
\begin{equation}
b(X_t) = a(X_t)-\frac{1}{2}\frac{\partial a(X_t)}{\partial X_t} b(X_t).
\label{secnon05}
\end{equation}
The second order continuous approximation for the map $X_t$ is then given by
\begin{equation}
\frac{dX_t}{dt} = \frac{a(X_t)}{1+\frac{1}{2}\frac{\partial a(X_t)}{\partial X_t} }.
\label{secnon06}
\end{equation}
The higher-order approximations are obtained by taking additional terms in the sum on the right hand side (r.h.s) of Eq.~(\ref{secnon04}) and exploiting the chain rule for the derivatives of $a(X_t)$. Instead of an algebraic equation for $b(X_t)$, we will obtain a differential equation. The differential equation for $b(X_t)$ is then transformed into an algebraic equation by substituting the explicit result of previous approximations for the derivatives 
$\partial b(X_t)/\partial X_t$.  For example, the third order expansion of $b(X_t)$ is provided by truncating the sum in Eq.~(\ref{secnon04}) at $B_{2k}=B_2$. 
$b(X_t)$  then satisfies the following equation 
$
b(X_t) = a(X_t)-\frac{1}{2}\frac{\partial a(X_t)}{\partial X_t} b(X_t)+\frac{B_2}{2}b(X_t)\frac{\partial}{\partial X_t}
\left(\frac{\partial a(X_t)}{\partial X_t} b(X_t) \right)
$. By means of Eq.~(\ref{secnon05}) , this is transformed into a quadratic equation for $b(X_t)$
\begin{equation}
\frac{B_2}{2}\frac{\partial^2 a(X_t)}{\partial X_t^2} b(X_t)^2
+\left[
\frac{B_2}{2}\frac{\partial}{\partial X_t} \left( \frac{a(X_t)}{1+\frac{1}{2}\frac{\partial a(X_t)}{\partial X_t}}\right)
-\frac{1}{2}\frac{\partial a(X_t)}{\partial X_t}-1
\right]b(X_t)+a(X_t)=0.
\label{secnon07}
\end{equation}
\begin{figure}[t]
\begin{center}
	\begin{subfigure}[b]{0.51\textwidth}
                \includegraphics[width=\textwidth]{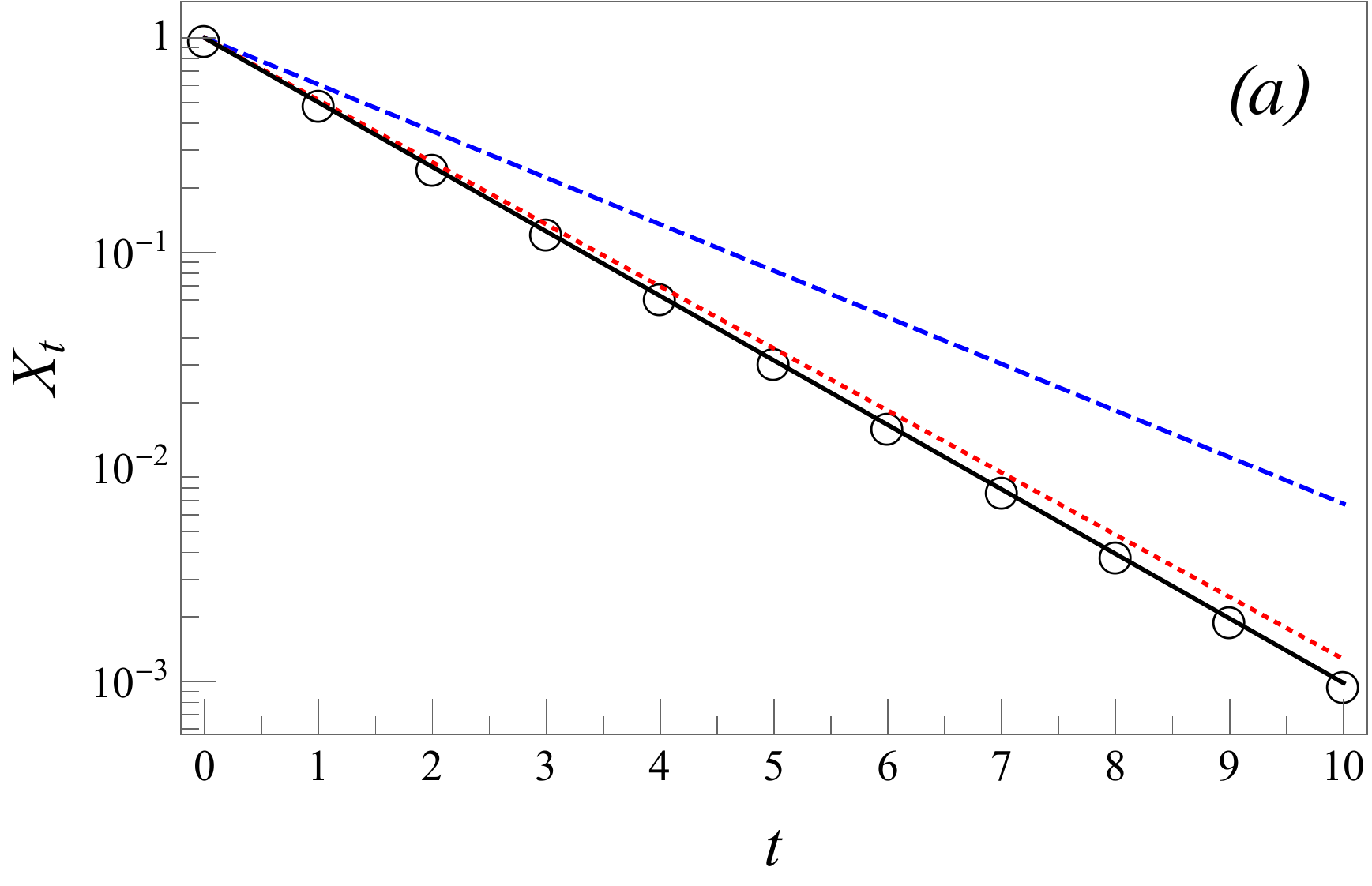}
              
        \end{subfigure}%
        ~
        \begin{subfigure}[b]{0.5\textwidth}
                \includegraphics[width=\textwidth]{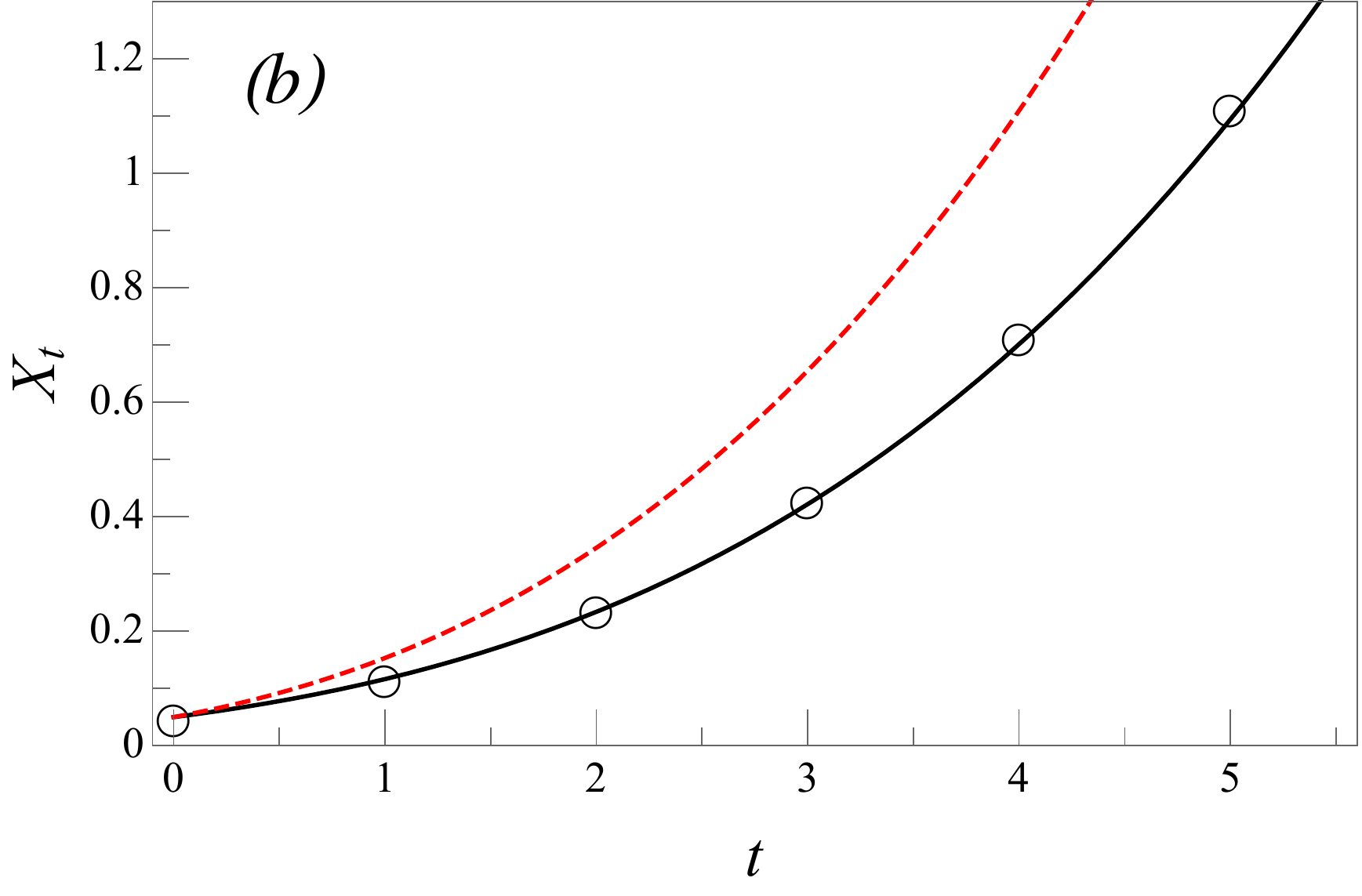}
              
        \end{subfigure}%

\end{center}
\caption{
Deterministic maps and their continuous approximations. In panel {\bf (a) } the dynamics of $X_t$ under the linear map $X_{t+1}=X_t-0.5 X_t$  ($X_0=1$) is plotted (circles) and 
the appropriate first (dashed), second (dotted) and third (thick) approximations. Panel {\bf (b)} describes the dynamics of the Pomeau-Manneville map Eq.~(\ref{pomeau}) (circles) and appropriate first (dashed) and second (thick) order approximations. The parameters of the Pomeau-Manneville map are $a_0=-0.5$, $\alpha=1.5$ and $X_0=0.05$. 
}
\label{fignonoise}
\end{figure}

The linear map of the form $X_{t+1}=X_t-\alpha X_t$ is a trivial example. The continuous approximation is always of the form 
$X_t=X_0\exp(-z_l(\alpha)t)$. The first, second and third order approximations are $z_1(\alpha)=\alpha$, $z_2(\alpha)=\alpha/(1-(1/2)\alpha)$ and $z_3(\alpha)=\alpha/(1-(1/2)\alpha-(1/12)\frac{\alpha^2}{1-(1/2)\alpha})$. The comparison to the solution 
$X_t=X_0(1-\alpha)^t$ is presented in Fig.~\ref{fignonoise} (a). It is clear that for such a simple case, all the approximations display the correct functional behavior of the solution, while the differences are in the order of  approximation of the decay constant $\log(1-\alpha)^{-1}$.

The Pomeau-Manneville map~\cite{PomeauManneville} is used as a model of intermittent behavior~\cite{Korabel},  given by the formula
\begin{equation}
X_{t+1}=X_t-a_0 X_t^{1/\alpha},  
\label{pomeau}
\end{equation}
where $X$  is unbounded. 
First and second order approximations are solutions of  $dX_t/dt=-a_0 X_t^{1/\alpha}$ and 
$dX_t/dt=-a_0 X_t^{1/\alpha}/\left(1-(1/2\alpha) a_0 X_t^{1/\alpha-1}\right)$, respectively. The solutions of these two differential equations are clearly different. In Fig.~\ref{fignonoise}(b) a numerical comparison between the actual behavior of the map  and these approximations is presented. From the figure it is clear that the second order and first order approximations are comparable 
in the vicinity of $t=0$. While the first order approximation strongly deviates from actual solution, the second order approximation stays very close for the whole domain $0\leq X_t \leq 1$.   

\section{First and Second Order Stochastic Map Approximations}
\label{stoch}

The treatment of a SM is similar in some sense to the approximation for the non-stochastic map. The equation for the SM is taken to be
\begin{equation}
X_{t+1}=X_t+a(X_t)+\eta_t,
\label{SM01}
\end{equation}
where $\eta_t$  is the stochastic part.
For simplicity we assume that $\eta_t$ is a random variable with Gaussian distribution, 
zero mean and constant second moment. We assume that the continuous approximation can be written in the following form
\begin{equation}
X_{t}=X_{t_0}+\int_{t_0}^{t} b(X_{t'})\,dt'+\int_{t_0}^{t}c(X_{t'})dB_{t'}
\label{SM02}
\end{equation}
for any $t$ and $t_0$. $b(X_{t})$ and $c(X_{t})$ are some as yet unknown functions of the random variable $X_t$. 
$B_t$ is a Wiener process, or physically speaking, the noise term. The differential analog of Eq.~(\ref{SM02}) is 
\begin{equation}
dX_t = b(X_t)\,dt+c(X_t)\,dB_t,
\label{SM03}
\end{equation}
which is similar, as mentioned above, to the Langevin equation with drift term, $b(X_t)$, and multiplicative noise, $c(X_t)$.
Similarly to what has been done in Sec.~\ref{nonstoch}, we impose that $\forall t_0$, $b(X_t)$ and $c(X_t)$ must satisfy:  
\begin{equation}
a(X_{t_0})+\langle \eta^2 \rangle \int_{t_0}^{t_0+1}\,dB_{t}= \int_{t_0}^{t_0+1}b(X_{t})\,dt+\int_{t_0}^{t_0+1}c(X_{t})\,dB_{t}.
\label{SMR}
\end{equation}
While the  first term on the  r.h.s is similar to the pure deterministic term in Eq.~(\ref{secnon02}), the next term on the r.h.s of Eq.~(\ref{SMR}) appears only due to the presence of stochastic term $\eta_t$ in SM. Having  assumed that $\eta_t$ is Gaussian and uncorrelated noise,  this tern can be decomposed into a sum of Gaussian variables,  which we approximate as $\int_{t_0}^{t_0+1}\,dB_{t}$ on the l.h.s of Eq.~(\ref{SMR}). 
When the noise is non-Gaussian or correlated, the representation of the noise as an integral over a Wiener process is not possible and the approximation might break down. The integral over $c(X_t)$ on the r.h.s is dictated by the general form of Langevin equation with multiplicative noise.

The next step of approximation is to write $a(X_t)$ as an integral. In Sec.~\ref{nonstoch} we 
used the Euler-Maclaurin formula for this purpose, but now we are treating stochastic variables and the usual 
rules of calculus do not apply. We start the approximation by stating that for the function $K(t) = (t-t_0)-1/2$ a trivial relation holds 
\begin{equation}
d(a(X_t)K(t)) = a(X_t)\frac{dK(t)}{dt}\,dt + K(t)d(a(X_t)).
\label{SM04}
\end{equation}
Eq.~(\ref{SM04}) looks similar to the usual calculus differential rules, but it is actually a specific case of a more 
general differential formula which involves Wiener processes. Specifically, 
\begin{equation}
d(a(X_t)) = \frac{\partial a(X_t)}{\partial X_t} \left( b(X_t)\,dt+c(X_t)\,dB_t \right) + 
\frac{1}{2}\frac{\partial ^2 a(X_t)}{\partial X_t^2}c(X_t)^2\,dt
\label{SM05}
\end{equation}
according to the It\^{o} formula~\cite{Gardiner}. By integration of Eq.~(\ref{SM04}) we obtain
\begin{equation}
a(X_{t_0}) = \int_{t_0}^{t_0+1} a(X_t)\,dt -\frac{1}{2}\int_{t_0}^{t_0+1} d(a(X_t)) + 
\int_{t_0}^{t_0+1} K(t)\,d(a(X_t)).
\label{SM06}
\end{equation}
The first term on the r.h.s gives the first order approximation, similar to the result that we obtained for the deterministic map. 
Here, $\int_{t_0}^{t_0+1} a(X_t)\,dt+\langle \eta^2 \rangle \int_{t_0}^{t_t+1}\,dB_{t}= \int_{t_0}^{t_0+1}b(X_{t})\,dt+\int_{t_0}^{t_0+1}c(X_{t})\,dB_{t}$ and the first order approximation is 
\begin{equation}
b(X_t) = a(X_t);\quad c(X_t) = \sqrt{\langle \eta ^2 \rangle}.
\label{SM07}
\end{equation}
The second order approximation includes the first and the second terms on the r.h.s of Eq.~(\ref{SM06}), i.e. 
$\int_{t_0}^{t_0+1} a(X_t)\,dt-1/2\int_{t_0}^{t_0+1} da(X_t)+\sqrt{\langle \eta^2 \rangle} \int_{t_0}^{t_t+1}\,dB_{t}= \int_{t_0}^{t_t+1}b(X_{t})\,dt+\int_{t_0}^{t_0+1}c(X_{t})\,dB_{t}$. 
By applying Eq.~(\ref{SM05}) we obtain 

\begin{equation}
\begin{aligned}
&\int_{t_0}^{t_0+1} \left( a(X_t) -\frac{1}{2}\frac{\partial a(X_t)}{\partial X_t} b(X_t)-
\frac{1}{4}\frac{\partial ^2 a(X_t)}{\partial X_t^2}c(X_t)^2\right)\,dt+
\int_{t_0}^{t_0+1}\left(
-\frac{1}{2}\frac{\partial a(X_t)}{\partial X_t} c(X_t)
+\sqrt{\langle \eta^2 \rangle }
\right)\,dB_{t}
\\
&
= 
 \int_{t_0}^{t_0+1}b(X_{t})\,dt+\int_{t_0}^{t_0+1}c(X_{t})\,dB_{t}.
\end{aligned}
\label{SM08}
\end{equation}
Comparison of the integrands provides the result 
\begin{equation}
\label{SM09}
b(X_t)  = \frac{a(X_t)-\frac{1}{4}\frac{\partial^2a(X_t)}{\partial X_t^2}\left( \frac{\langle \eta^2 \rangle}{1+\frac{1}{2}\frac{\partial a(X_t)}{\partial X_t}}\right)^2}{1+\frac{1}{2}\frac{\partial a(X_t)}{\partial X_t}};
\quad
c(X_t) = \frac{\sqrt{\langle \eta^2 \rangle}}{1+\frac{1}{2}\frac{\partial a(X_t)}{\partial X_t}},
\end{equation}
and the second-order continuous approximation of the SM is 
\begin{equation}
\label{SM10}
dX_t = \frac{a(X_t)-\frac{1}{4}\frac{\partial^2a(X_t)}{\partial X_t^2}\left( \frac{\sqrt{\langle \eta^2 \rangle}}{1+\frac{1}{2}\frac{\partial a(X_t)}{\partial X_t}}\right)^2}{1+\frac{1}{2}\frac{\partial a(X_t)}{\partial X_t}}\,dt + 
\frac{\sqrt{\langle \eta^2 \rangle}}{1+\frac{1}{2}\frac{\partial a(X_t)}{\partial X_t}}\,dB_t.
\end{equation}
Eq.~(\ref{SM10}) describes a stochastic process for which both the drift term and the noise term depend on $X_t$. Noise which is dependent not only on time but also on the coordinate is termed a multiplicative noise~\cite{Gardiner,GitBook,GitBur,Bena,Lau}. From Eq.~(\ref{SM10}) it becomes clear that the behavior of the SM is quite different from the behavior described by a Langevin equation with thermal noise.
The next terms of the expansion are derived from the expansion of the remainder $\int_{t_0}^{t_0+1} K(t)\,d(a(X_t))$ and are not treated in this manuscript.

While  here we have developed a  continuous SDE approximation of the SM, several methods exist for discretization of SDEs~\cite{Kloeden}, turning them into SMs. In most cases the SDEs are not amenable to an analytical solution and a numerical approach is used.  Various integration schemes for SDEs such as 
Euler-Maruyama~\cite{Kloeden}, Milstein~\cite{Milstein}, Stochastic Runge-Kutta~\cite{Kloeden}, Local-Linearization~\cite{Ozaki1,Ozaki2,Biscay}, have been developed. One  feature of these schemes is that an unfortunate choice of the discretization parameter (e.g. $\Delta t$) can lead to unbounded behavior of the solution, basically unstable behavior. In Sec.~\ref{stabilityC} we discuss the situation when the continuous approximation of an SM dictates an nonnormalizable stable distribution, an outcome of unstable behavior of the SM. This stability criterion for an SM can be applied towards integration schemes of SDEs. The  simple SM in Eq.~(\ref{SM01})  can be viewed as a simple integration scheme with $\Delta t=1$, by generalizing it to arbitrary $\Delta t$ and rederiving the continuous approximation the stability of the scheme can be tested by probing the normalizability of the stable distribution. 

 We now turn to a comparison between the behavior 
 of various SMs and their approximations, as given by Eq.~(\ref{SM10}). A single realization of the SM will provide a random 
 trajectory for $X_t$ for different $t$, as will Eq.~(\ref{SM10}) and Eq.~(\ref{SM07}). Instead of comparing different random trajectories, we will compare the approximations to the stable distributions of $X_t$ (if they exist) for $t\to\infty$.

\section{Stable Distributions}
\label{stable}

For $X_t$ described by an SM of a form similar to Eq.~(\ref{SM01}) the behavior is random due to the presence of the noise term $\eta(t)$. The distribution of $X_t$ changes in time and is given by $P_t(X_t)$.  For some classes of SMs this distribution will converge to a stationary distribution:  
$P_t(X_t)\to P(X)$ as $t\to\infty$, where $X$ now describes all possible values of the coordinate at long times.
The existence of a stable distribution and its shape is generally determined by  means of a numerical simulation 
of the stochastic process described by Eq.~(\ref{SM01}), with the single exception of a linear $a(X_t)$, in which case $P(X)$ can be determined with existing techniques in closed form~\cite{McKane2,Amir1,Naama1}. However, 
the continuous approximations of the previous section provide a route for the analytical calculation of an approximate 
stable distribution via the solution of the appropriate Fokker-Planck equation. In the following we provide 
the stationary solution of a Fokker-Planck equation for the process described by Eqs~(\ref{SM07}) and (\ref{SM10}); 
this is a standard task widely described in the literature~\cite{Risken}.

The appropriate Fokker-Planck equation for the stochastic process in Eq.~(\ref{SM03}) is 
\begin{equation}
\label{stable01}
\frac{\partial P_t(X_t)}{\partial t} = \frac{\partial}{\partial X_t}\left(-b(X_t)\right)P_t(X_t)
+\frac{1}{2}\frac{\partial^2 }{\partial X_t^2} \left( c(X_t)^2P_t(X_t) \right),
\end{equation}
where the It\^{o} convention was used~\cite{Lau}. In the limit $t \to \infty$, $X_t\to X$ and $P_t(X_t)\to P(X)$, Eq.~(\ref{stable01}) is transformed into 
\begin{equation}
\label{stable02}
 \frac{d}{d X}\left[\left(-b(X)+c(X)\frac{dc(X)}{dX}
+ \frac{1}{2}c(X)^2\frac{d}{dX} \right)P(X)\right] = 0.
\end{equation}
Eq.~(\ref{stable02}) can be compared to the standard form of the stable solution of a Fokker-Planck equation for a thermal process. The main difference is the presence of the term $c(X)dc(X)/dX$ which produces an additional drift due to the multiplicative nature of the noise. We look for a solution of the form $P(X)={\cal N}^{-1} \exp\left(-H(X)\right)$, where $\cal N$ is the normalization constant ${\cal N} = \int_{-\infty}^{\infty}\exp\left(-H(X)\right)\,dX$ . The function $H(X)$ satisfies the equation
\begin{equation}
\label{stable03}
\frac{dH(X)}{dX} = \frac{d\ln\left(c(X)^2\right)}{dX}-\frac{2b(X)}{c(X)^2}.
\end{equation}
From Eq.~(\ref{SM10}) we arrive at a simple form for $H(X)$ for the second-order approximation
\begin{equation}
\label{stable04}
H(X) = -\frac{2}{\eta^2}\int a(X)\,dX-\frac{1}{2\eta^2}a(X)^2-\ln\left(\left|1+\frac{1}{2}\frac{da(X)}{dX}\right| \right).
\end{equation}
For any given approximation of a SM we can now write the stable distribution of $X$, given it exists.
We now show a few examples.

\begin{figure}[t]
\begin{center}
	\begin{subfigure}[b]{0.51\textwidth}
                \includegraphics[width=\textwidth]{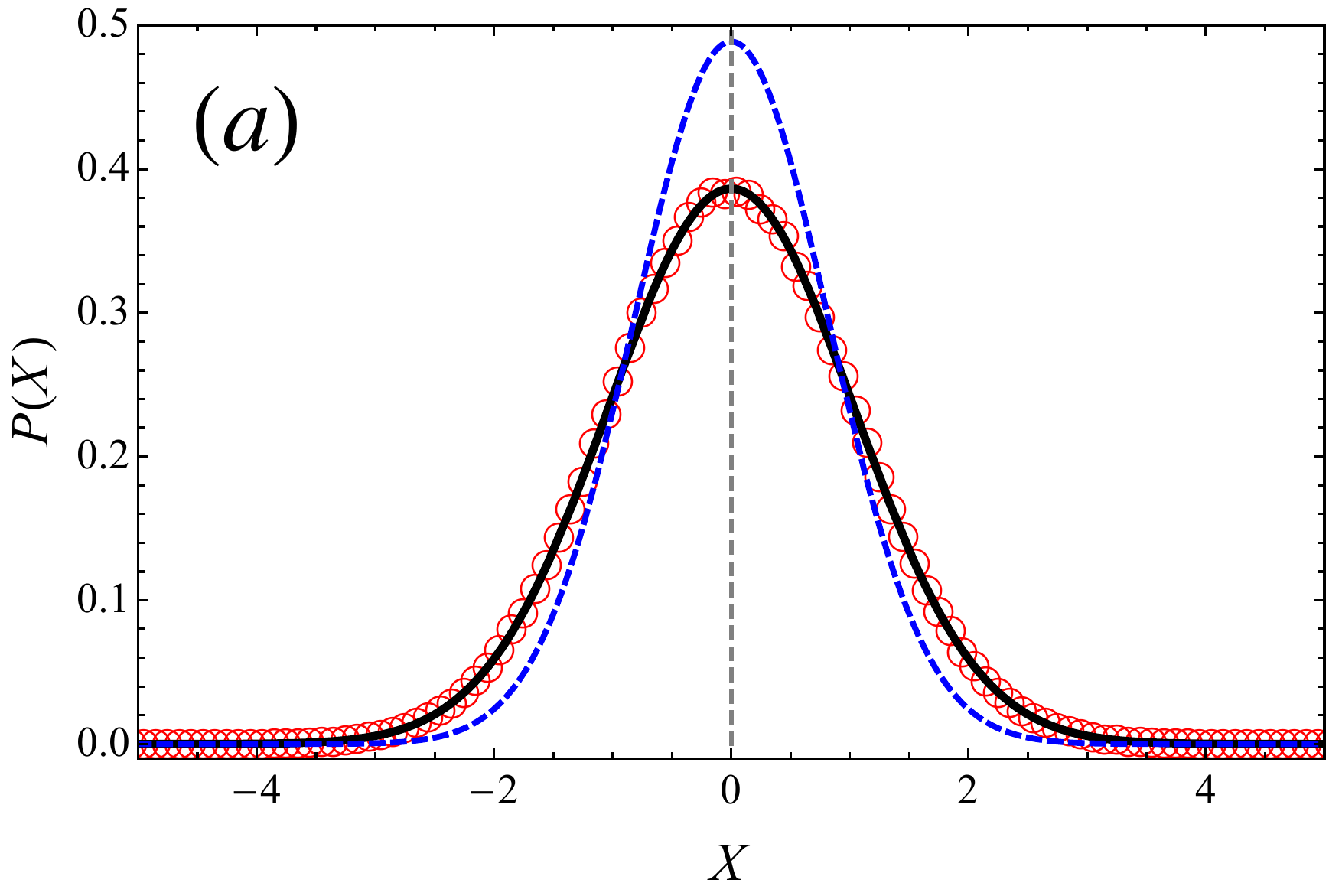}
              
        \end{subfigure}%
        ~
        \begin{subfigure}[b]{0.51\textwidth}
                \includegraphics[width=\textwidth]{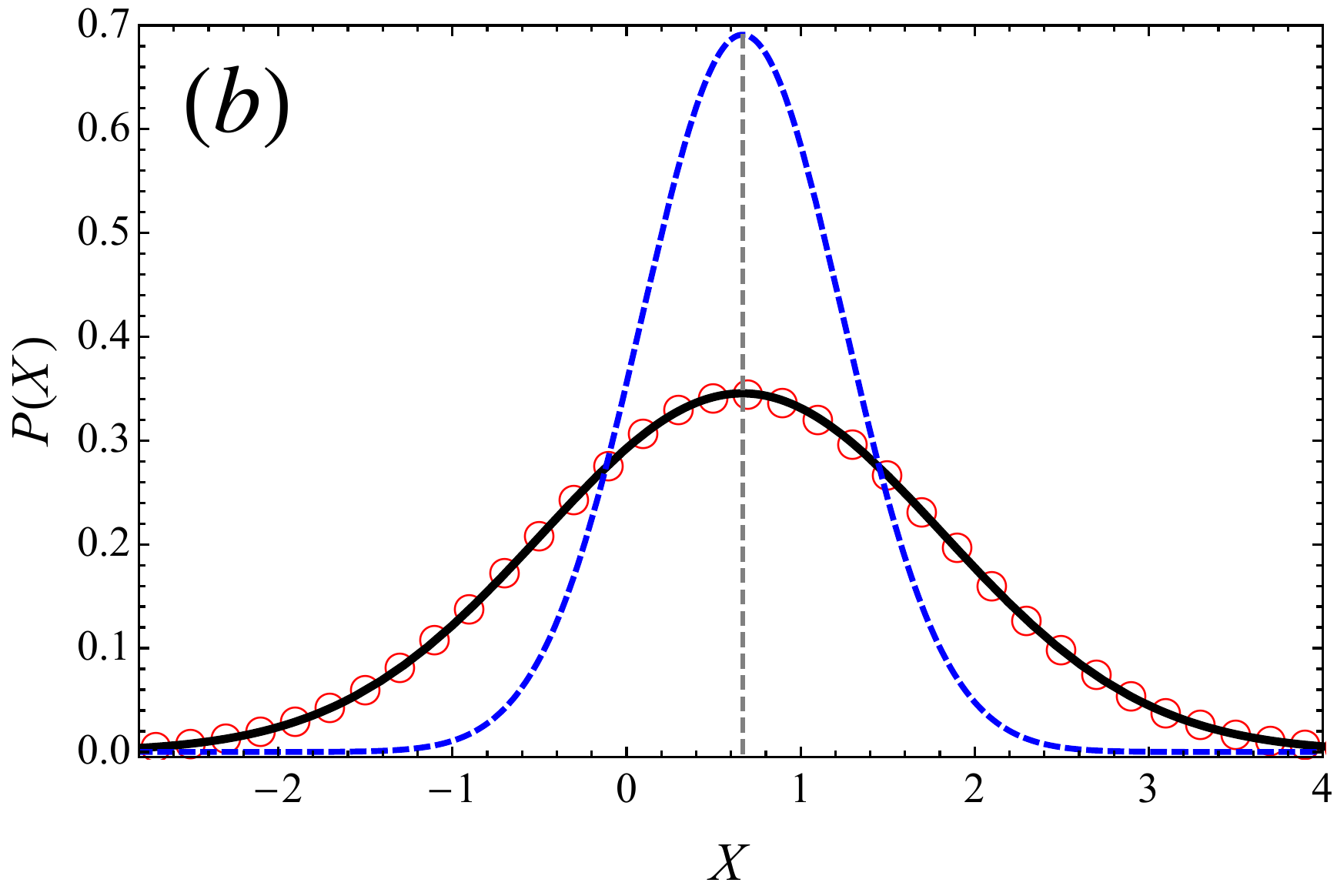}
              
        \end{subfigure}%

\end{center}
\caption{
Stable distributions of linear stochastic maps. In panel {\bf (a) } the linear map $X_{t+1}=X_t-\alpha X_t+\eta_t$ is plotted (circles), the noise term is zero mean, a Gaussian and $\langle \eta_t^2 \rangle=1$, $\alpha=0.75$. The appropriate first-order (dashed line) and second-order (thick line) approximations are plotted. Panel {\bf (b)} is similar to panel  {\bf (a)}, except $\alpha=1.5$, $\langle \eta_t \rangle =1$ and $\langle \eta^2 \rangle = 2$. 
}
\label{figlin}
\end{figure}

\subsection{Linear Map}

The linear SM is the stochastic version of the map described at the end of Sec.~\ref{nonstoch}, i.e. $a(X_t) = -\alpha X_t$, 
with constant $\alpha$. The noise term in Eq.~(\ref{SM01}) is assumed to be Gaussian with zero mean and constant 
second moment ($\langle \eta_t^2 \rangle = \langle \eta^2 \rangle$),
\begin{equation}
\label{linmap01}
X_{t+1}= X_t -\alpha X_t + \eta_t.
\end{equation}
According to Eq.~(\ref{SM07}) the first-order SM approximation is 
\begin{equation}
\label{linmap02}
dX_t = -\alpha X_t\,dt+\sqrt{\langle \eta^2\rangle}\, dW_t, 
\end{equation}
and therefore, according to Eq.~(\ref{stable03}), the corresponding first-order stable distribution $P_{1}(X)$ is 
\begin{equation}
\label{linmap03}
P_{1}(X) =\sqrt{\frac{\alpha}{\pi\langle \eta^2 \rangle}} \exp\left( -\frac{\alpha X^2}{\langle \eta^2 \rangle}\right).
\end{equation}
This distribution is Gaussian and is compared to the numerical result in Fig.~\ref{figlin}, showing significant deviations. The second moment of the first-order approximation is off by a factor of $2$ (see Eq.~(\ref{linmap05}) which is exact at $\alpha=1$). 
Moreover, the distribution $P^{1}(X)$ does not significantly change as a function of $\alpha$, while for $\alpha\geq2$ there is no stable distribution since in the non-stochastic form $|X_t|$ will attain larger and larger values as a function of $t$. 

The second-order approximation, Eq.~(\ref{SM10}), for the SM Eq.~(\ref{linmap01}).  yields the following 
continuous form
\begin{equation}
\label{linmap04}
dX_t = -\frac{\alpha}{1-\frac{1}{2}\alpha} X_t\,dt+\frac{\sqrt{\langle \eta^2\rangle}}{1-\frac{1}{2}\alpha}\, dW_t, 
\end{equation}
with the corresponding stable distribution $P_2(X)$ of the form
\begin{equation}
\label{linmap05}
P_{2}(X) =\sqrt{\frac{\alpha-\frac{1}{2}\alpha^2}{\pi\langle \eta^2 \rangle}} \exp\left( -\frac{\left(\alpha-\frac{1}{2}\alpha^2\right) X^2}{\langle \eta^2 \rangle}\right).
\end{equation}
The comparison to numerics now shows a perfect fit and, moreover, $P_{2}(X)$ is the exact stable distribution $P(X)$ for 
the linear map in Eq.~(\ref{linmap01})~\cite{McKane2,Amir1,Naama1}. It is somewhat surprising that while the non-stochastic second order approximation 
for the same map produces only an approximation, the stable distribution for a linear SM coincides precisely with the second-order 
approximation. The nonexistence of a stable distribution for $\alpha\ge 2$ is signaled by the fact that for $\alpha\to 2$ the second moment 
of $P_{2}(X)$ diverges. 

When the linear map is shifted by some constant $a_0$, i.e., $a(X) = -\alpha X+a_0$, the form of the solution does not change much. The normalization constant changes and the function is still a Gaussian, but with non-zero mean, 
$\sim\exp\left( -(\alpha-\alpha^2/2)(X-a_0/\alpha)^2/\langle \eta^2\rangle\right)$. The same effect occurs when the noise has a non-zero mean, $\langle \eta_t \rangle=a_0$, as presented in Fig.~\ref{figlin}.


\begin{figure}[t]
\begin{center}
	\begin{subfigure}[b]{0.51\textwidth}
                \includegraphics[width=\textwidth]{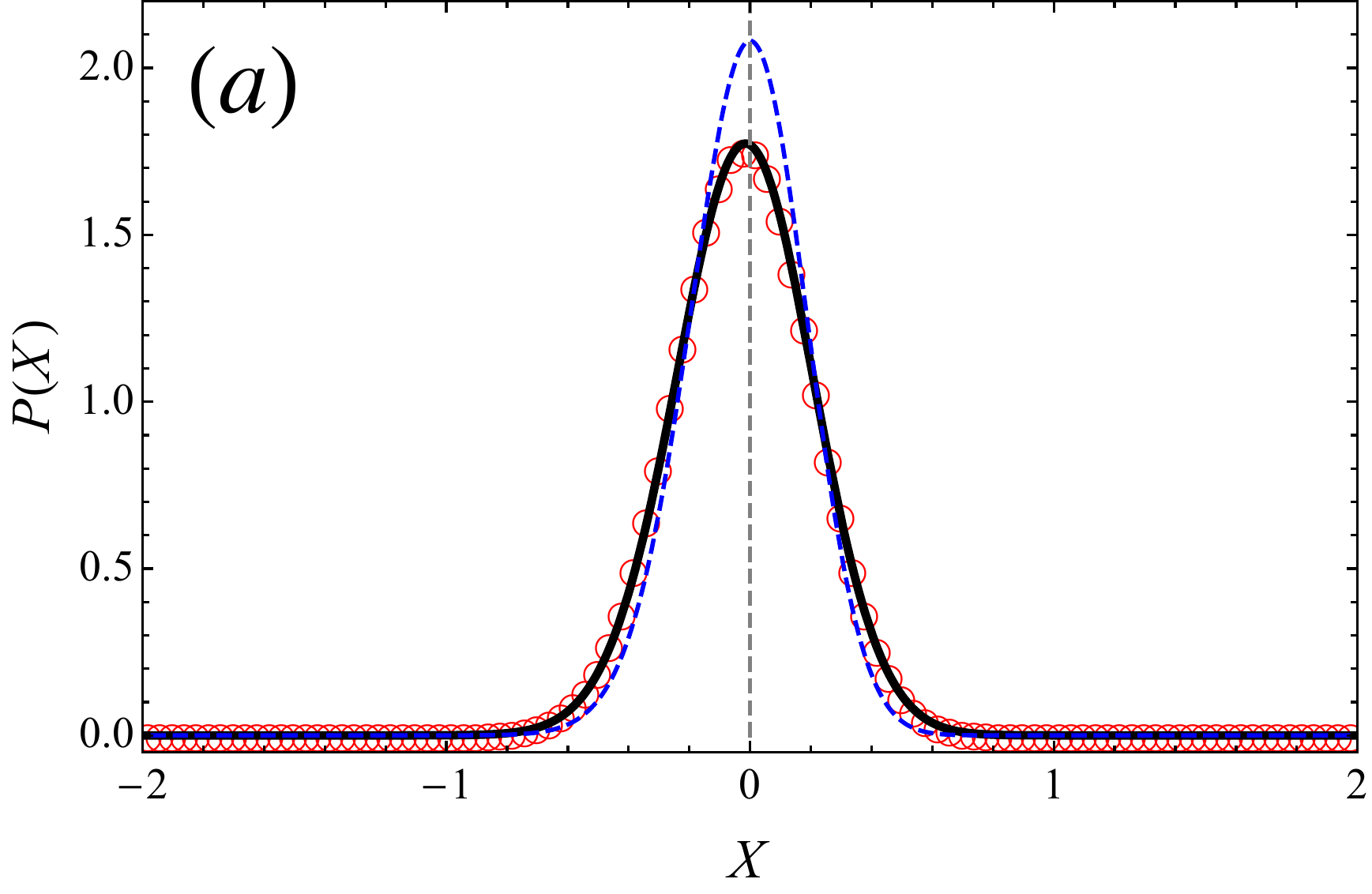}
              
        \end{subfigure}%
        ~
        \begin{subfigure}[b]{0.51\textwidth}
                \includegraphics[width=\textwidth]{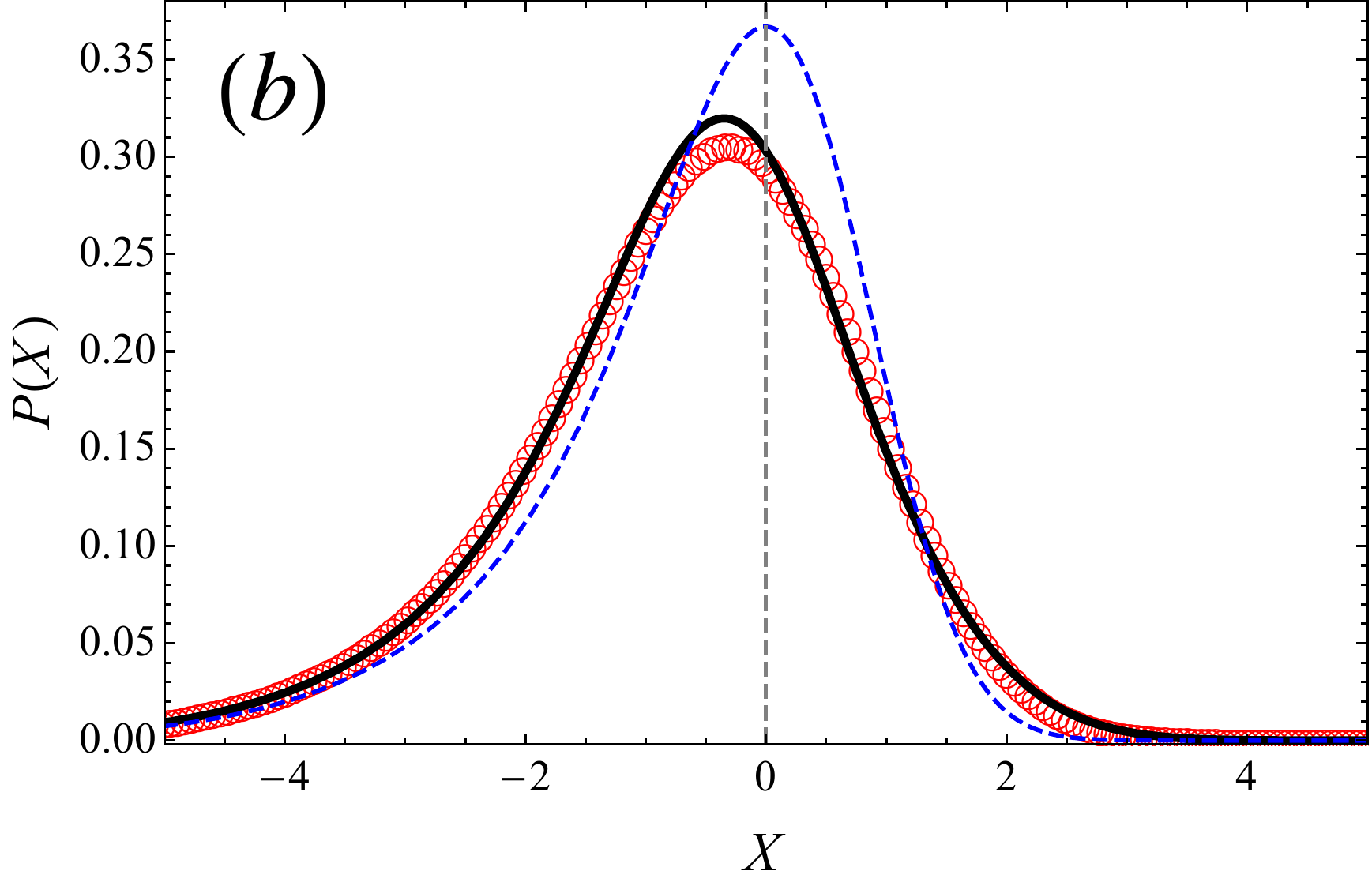}
              
        \end{subfigure}%

\end{center}
\caption{
Stable distributions of piece-wise linear stochastic maps. In panel {\bf (a) } the mapping given by Eq.~(\ref{expo01}) is plotted (circles), the noise term is zero mean, Gaussian and $\langle \eta_t^2 \rangle=0.2^2$, $p_-=1$ and $p_+=0.1$. The appropriate first order (dashed line) and second order (thick line) approximations are plotted. Panel {\bf (b)} is similar to panel  {\bf (a)}, except  $\langle \eta_t^2 \rangle =1$. 
}
\label{figpiece}
\end{figure}

\subsection{Non-Linear Maps}
\label{nonlinSec}
\subsubsection{Asymptotically Linear Map}
\label{expoMap} 
Consider the behavior of the map
\begin{equation}
\label{expo01}
X_{t+1} = X_{t} -p_-\frac{X_t}{1+\exp\left(-X_t\right)}-p_+\frac{X_t}{1+\exp\left( X_t \right)}+\eta_t.
\end{equation}
The function $a(X_t)$ of the presented map is asymptotically ($X_t\to\pm\infty$) linear with coefficients $p_+$ and $p_-$.  
The function $H(X)$ for the stable distribution of $X_t$, $P(X)\sim\exp\left(-H(X)\right)$ is  
\begin{equation}
\label{expo02}
\begin{aligned}
H(X) =&
-\frac{2}{\langle \eta^2 \rangle}\left\{ -\frac{p_-}{2}X^2+(p_- - p_+)\left[X\ln\left(1+\exp(X)\right)+Li_{2}\left(\exp(X)\right)\right]\right\}
\\&
-\frac{1}{2\langle \eta^2 \rangle}\left\{
\left[p_-\frac{X_t}{1+\exp\left(-X_t\right)}p_+\frac{X_t}{1+\exp\left( X_t \right)}\right]^2
\right\}
\\&
-\ln\left(\left|1+\frac{1}{2}\left[ 
\frac{p_-\left(\exp(X)(X-1)-1\right)-p_+\left( 1+X+\exp(X)\right)\exp(X)}
{\left(1+\exp(X)\right)^2}
\right]\right| \right),
\end{aligned}
\end{equation}
according to Eq.~(\ref{stable04}). $Li_n(z)=\sum_{k=1}^{\infty}z^k / k^n$ is the Polylogarithm function~\cite{Erdely}.
Each of the terms on the r.h.s of Eq.~(\ref{expo02}) corresponds to a term on the r.h.s. of 
Eq.~(\ref{stable04}). The first term on the r.h.s of Eq.~(\ref{expo02}), i.e. $\int a(X) \,dX$, corresponds to the first order approximation while the other two terms are the corrections due to the second-order scheme. The form of $H(X)$ in Eq.~(\ref{expo02}) is quite interesting. Specifically, we notice that $H(X)$ is not simply proportional to $1/\langle \eta^2 \rangle$. This means that increasing the noise strength will not simply lead to rescaling as one would expect for the Langevin description (with additive noise). In Fig.~\ref{figpiece}{\bf(a)} we plot the behavior of $P(X)$ for a piece-wise linear map ($p_-=1$ and $p_+=0.1$) and very low noise $\langle \eta^2 \rangle=0.2^2$. The maximum of the distribution is located at $X=0$ as is expected from the deterministic stable point $\alpha(0)=0$. No other deterministic fixed points exists.  In panel {\bf(b)} we increase the noise while leaving the deterministic parameters unchanged. We notice that the maximum of the distribution has changed to $X\approx -0.35$.  The first order approximation of $H(X)$ is homogeneous in $\langle \eta ^2 \rangle$ and still has its maximum at $X=0$. The second order approximation (while slightly off the numerically obtained distribution) predicts this effect correctly. For the SM there is coupling between noise and the non-linearity (or asymmetry) of  the mapping. Due to this coupling one must be cautious while addressing the noise strength as an effective temperature, the noise has multiplicative properties and creates an effective force~\cite{Lau}.

\begin{figure}[t]
\begin{center}
\begin{subfigure}[b]{0.4255\textwidth}
                \includegraphics[width=\textwidth]{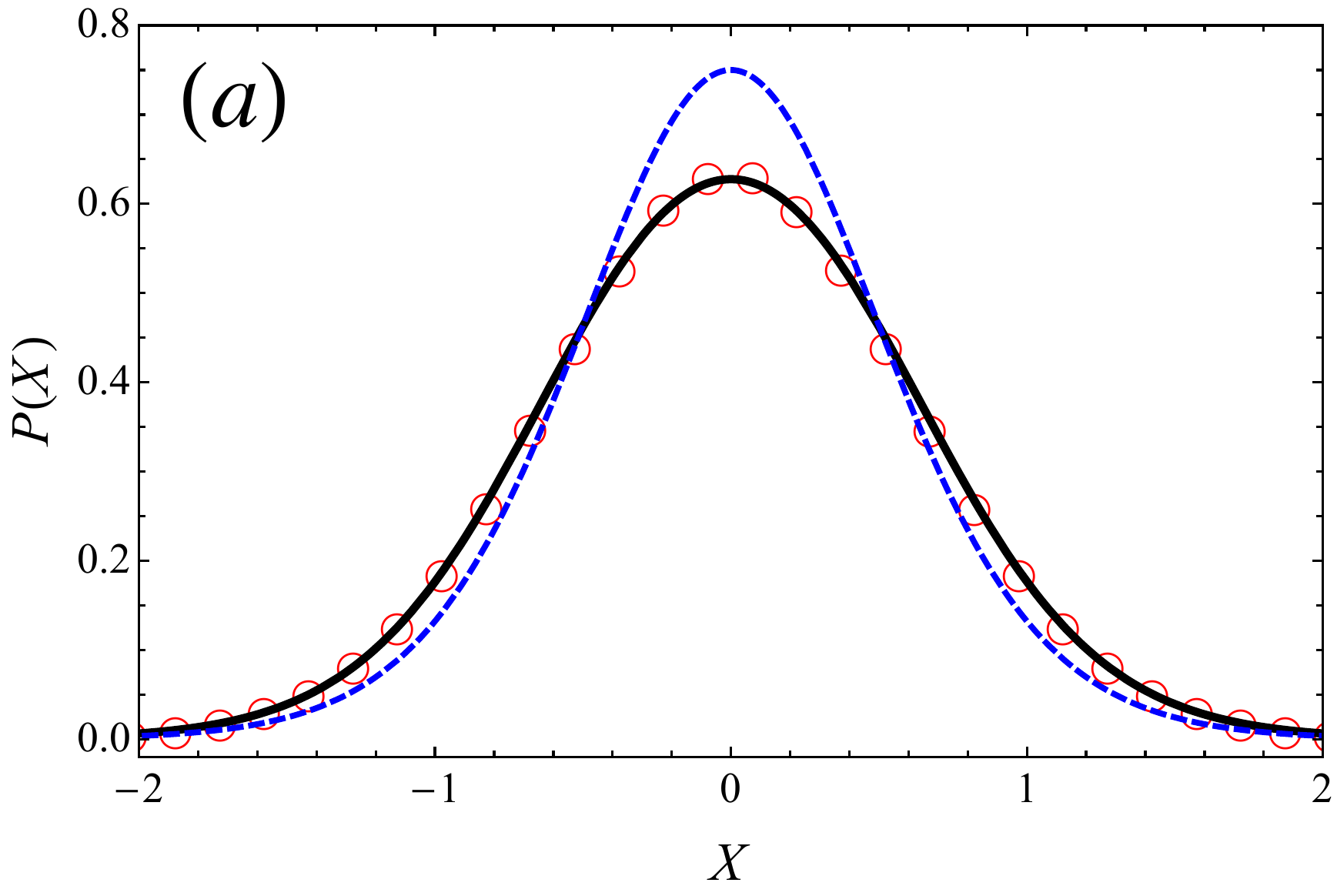}
              
        \end{subfigure} ~ 
\begin{subfigure}[b]{0.43255\textwidth}
                \includegraphics[width=\textwidth]{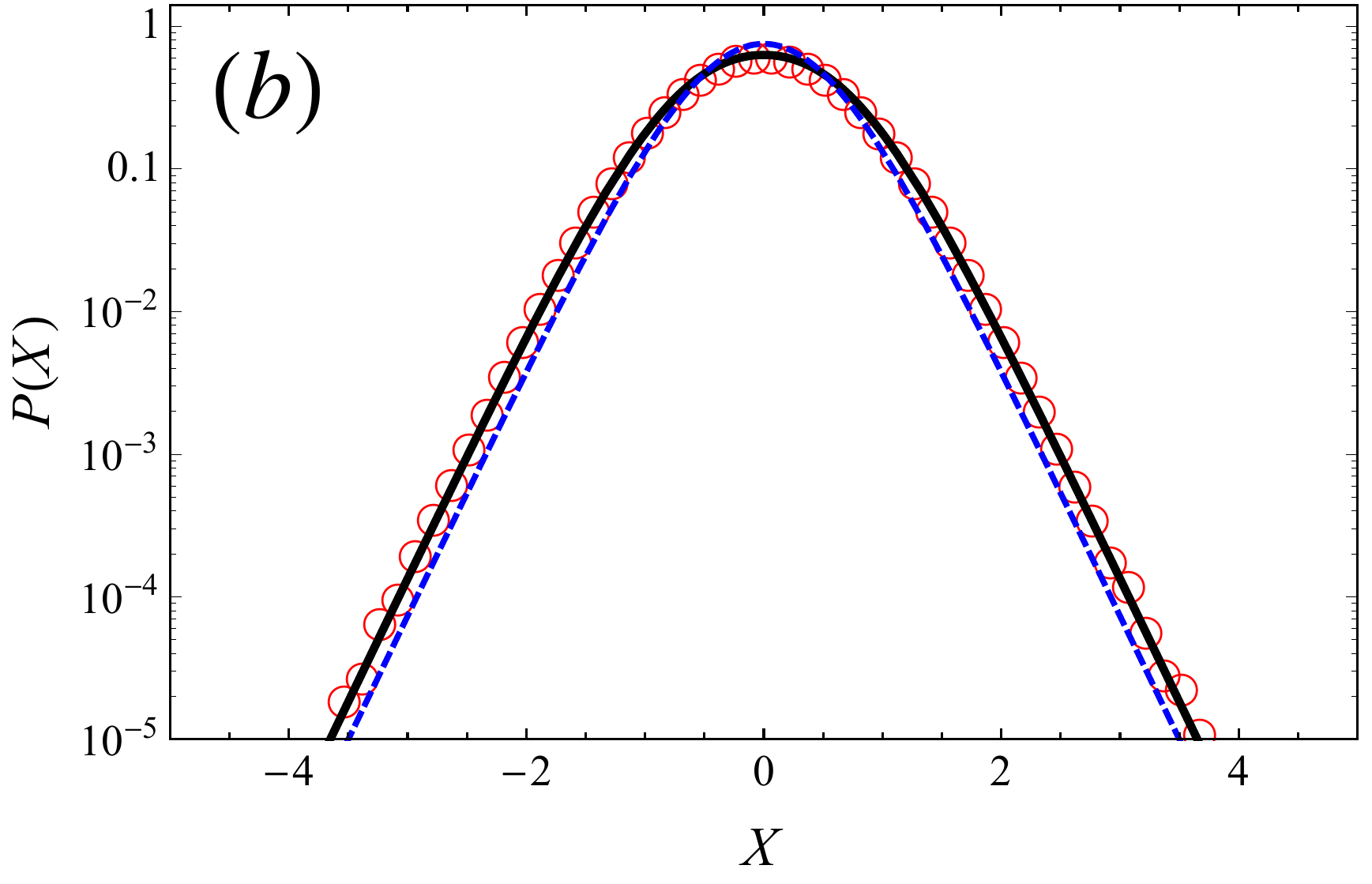}
              
        \end{subfigure} \\[0pt]
\begin{subfigure}[b]{0.4255\textwidth}
                \includegraphics[width=\textwidth]{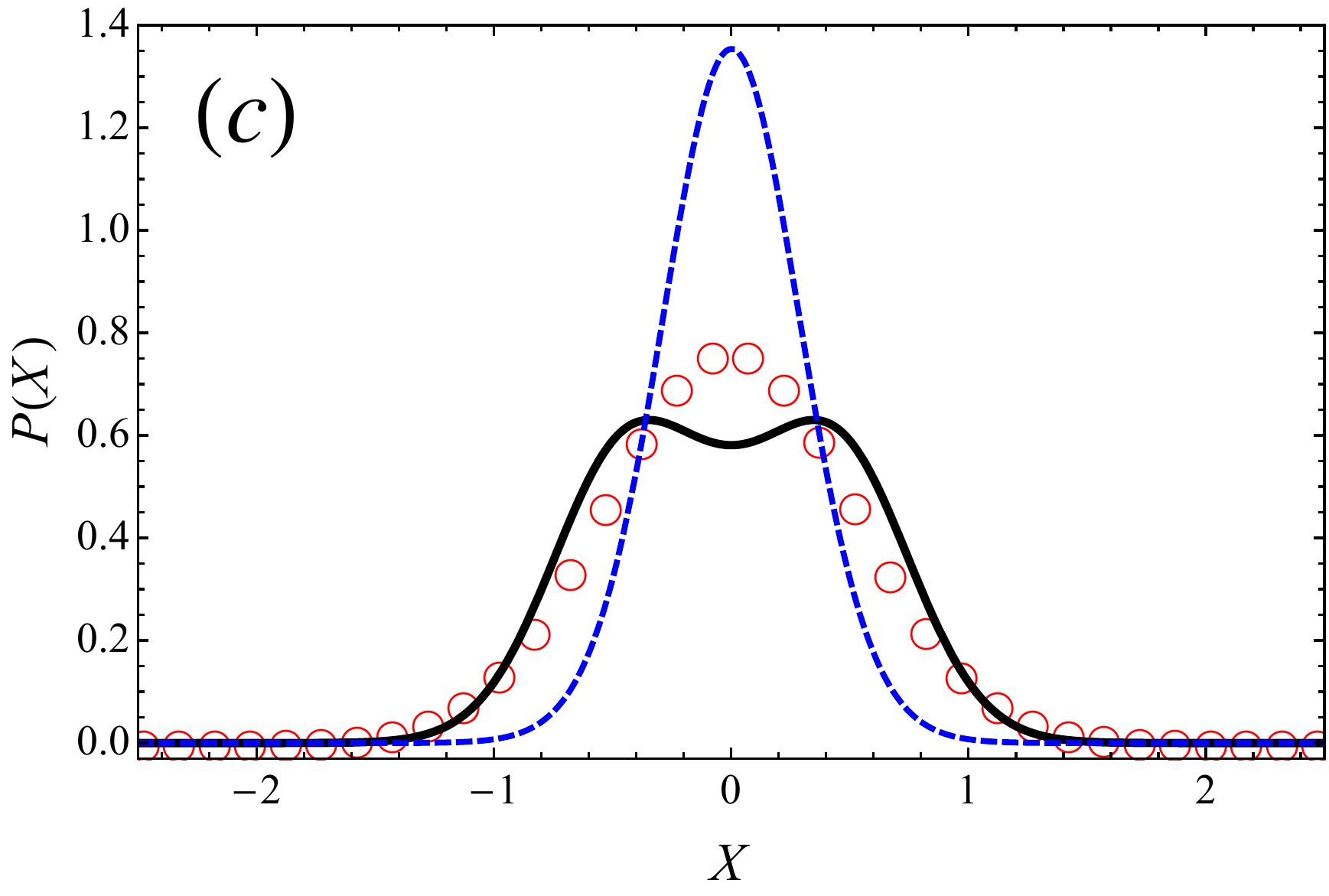}
              
        \end{subfigure} ~ 
\begin{subfigure}[b]{0.4255\textwidth}
                \includegraphics[width=\textwidth]{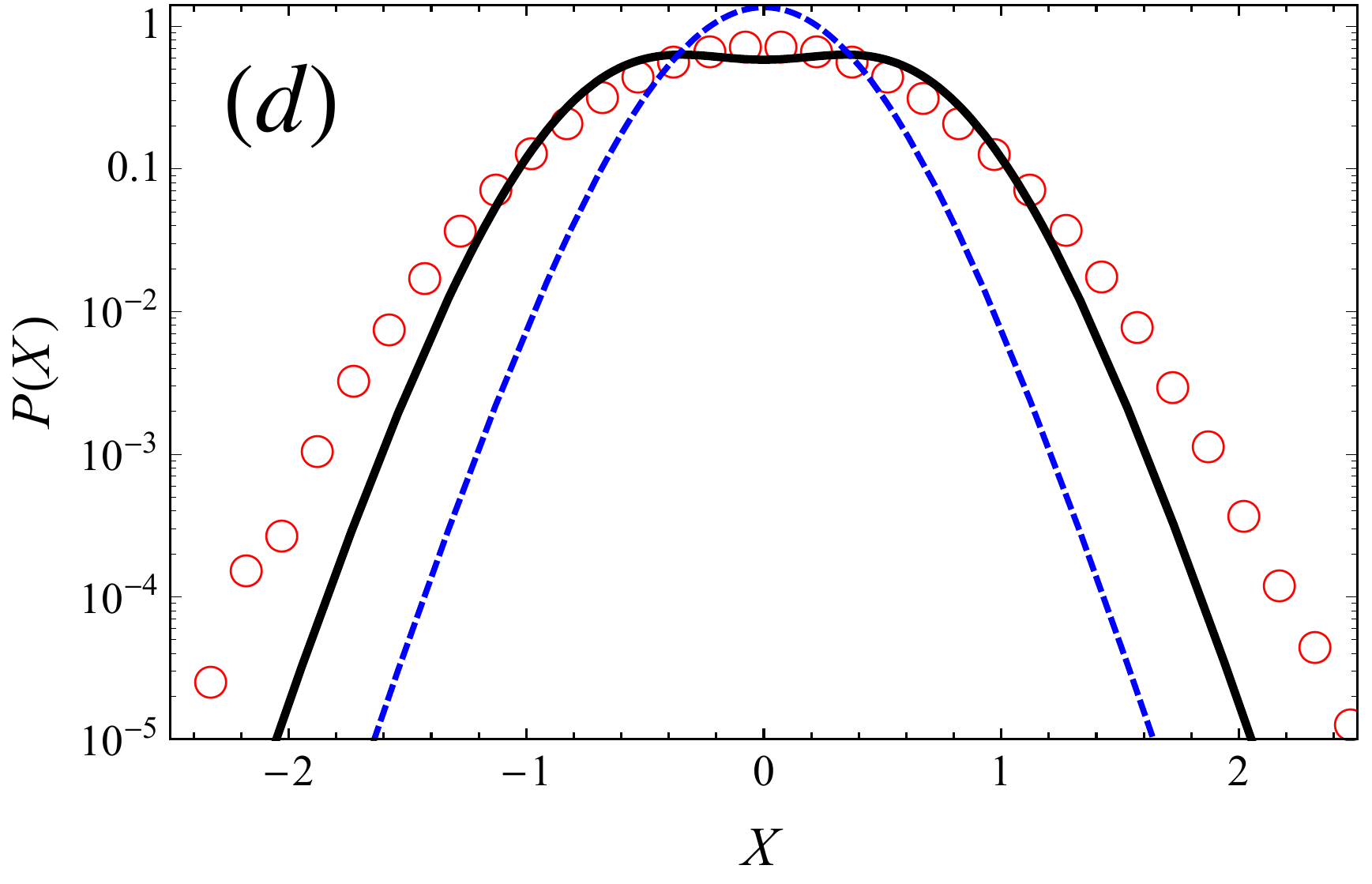}
              
        \end{subfigure}
\end{center}
\caption{ Stable distributions of stochastic maps with the $\tanh$ mapping, Eq. (\ref{hyp01}). In panel {\bf (a) } the stable distribution for the mapping  is plotted (circles), the noise term is zero mean, Gaussian and $\langle \eta_t^2 \rangle=0.5^2$, $z=0.5$. The  first order (dashed line) and second order (thick line) approximations are plotted. Panel {\bf (b)} is similar to panel  {\bf (a)} but plotted on a semi-log scale. The decay of the distribution is exponential. Panel {\bf (c)} is similar to panel  {\bf (a)} except $z=1.5$ and panel {\bf(d)} is similar to panel {\bf(c)} only plotted on a semi-log scale.
}
\label{fighyp}
\end{figure}

\subsubsection{Hyperbolic Tangent Map}
\label{HypMap} 

We define a non-linear map of the following form
\begin{equation}
\label{hyp01}
X_{t+1} = X_{t} -z \tanh \left(X_t\right)+\eta_t.
\end{equation}
The function $a(X_t)$ is asymptotically constant, $z$ is a parameter. The function $H(X)$ for the approximate stable distribution of $X_t$, $P(X)\sim\exp\left(-H(X)\right)$ is  
\begin{equation}
H(X) = -\frac{2}{\langle \eta^2 \rangle} z\ln\left[\cosh\left(X\right) \right]
-\frac{1}{2\langle \eta^2 \rangle}\left( z\tanh(X)\right)^2
-\ln\left[\left|1+\frac{1}{2} z\, \text{sech} (X)^2\right| \right].
\label{hyp02}
\end{equation}
In Fig.~\ref{fighyp} we present various behaviors of the stable distribution. For sufficiently low values of the noise strength and the parameter $z$ the fit is very good. The behavior is non-Gaussian since the decay of the stable distribution is exponential, as can be seen from panel {\bf(b)} of the figure. The decay follows $\sim\exp(-2zx/\langle \eta^2 \rangle)$. Panels {\bf(c)} and {\bf(d)} show that the technique developed in this manuscript is still only an approximation.  A discrepancy between the approximation and the actual behavior is observed as the parameter $z$ is increased. While the second-order approximation exhibits a double-peaked distribution, the simulation reveals a single maxima. The two peaks are a signature of the appearance of a limit cycle for the deterministic part of the mapping, but it wiped out, as seen in the simulation, by the presence of noise. For larger values of $z$ the mapping does start to show the presence of two phases (not shown) but the quality of the approximation in this regime is even worse.

\begin{figure}[t]
\begin{center}
\begin{subfigure}[b]{0.47\textwidth}
                \includegraphics[width=\textwidth]{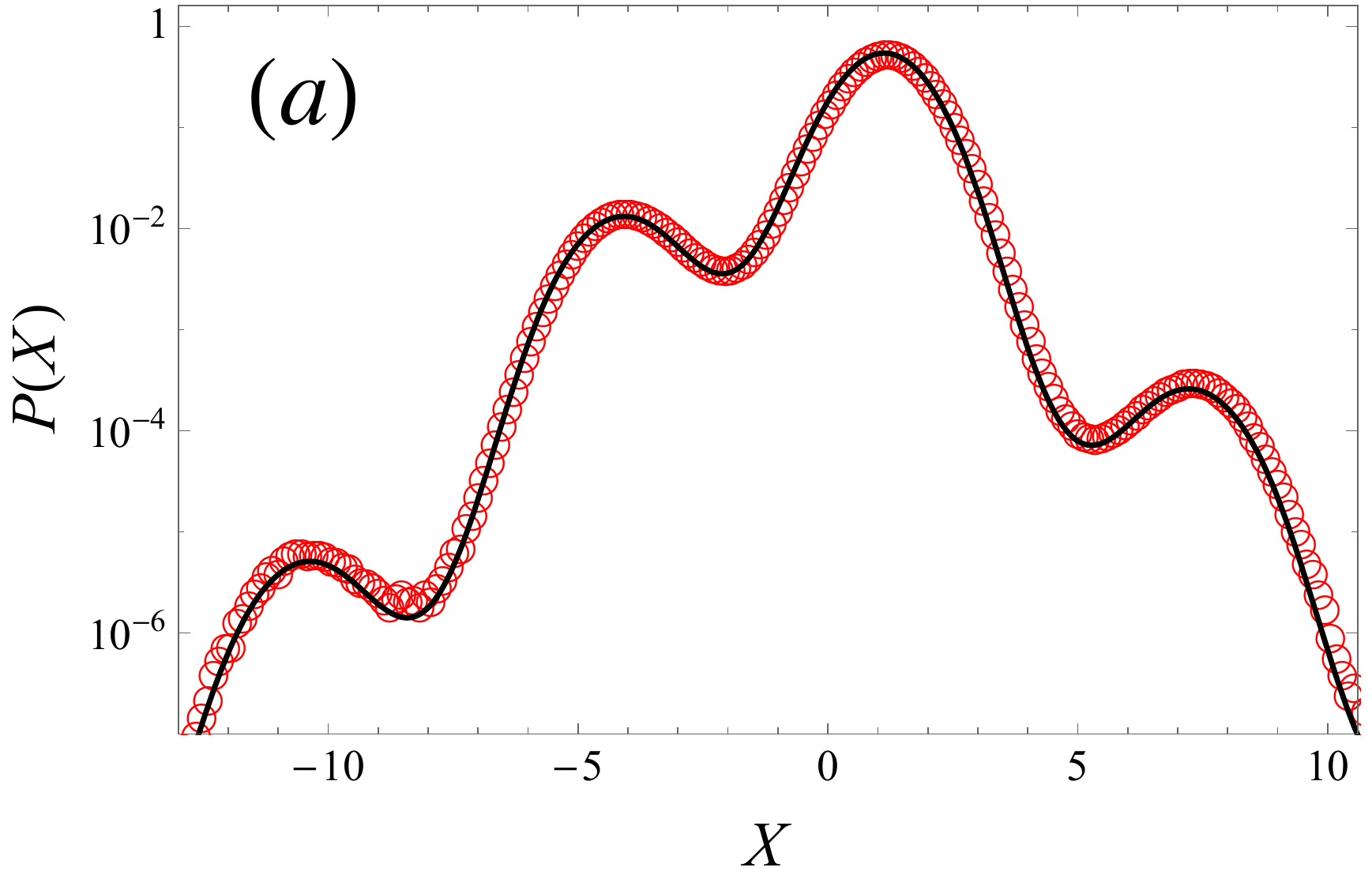}
              
        \end{subfigure} ~ 
\begin{subfigure}[b]{0.47\textwidth}
                \includegraphics[width=\textwidth]{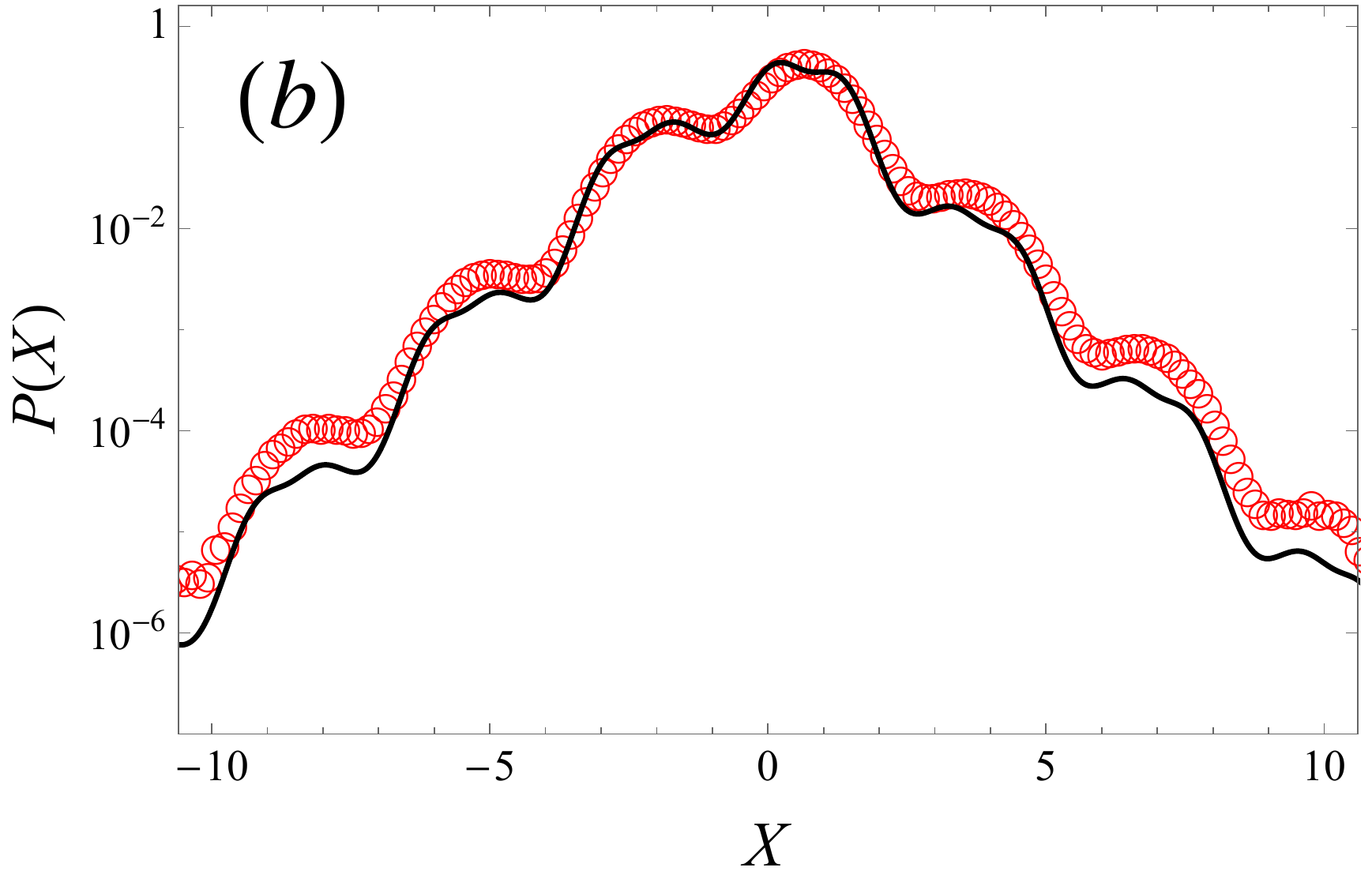}
              
        \end{subfigure} \\[0pt]
%
%
\end{center}
\caption{ Stable distributions of stochastic maps with oscillating mapping. Panel {\bf (a) } presents  the stable distribution for the map  $a(X)=-0.25\tanh(X)+0.5\cos(X)$  (circles), the noise term is zero mean, Gaussian and $\langle \eta_t^2 \rangle=0.4^2$. The  second order (thick line) approximation is plotted. In panel {\bf (b)} the map has higher local derivatives, i.e.  $a(X)=-0.25\tanh(X)+0.5\cos(2X)$  (circles), the noise term is zero mean, Gaussian and $\langle \eta_t^2 \rangle=0.4^2$. The  second order (thick line) approximation is plotted.}
\label{figoscill}
\end{figure}

\begin{figure}[t]
\begin{center}
\begin{subfigure}[b]{0.4255\textwidth}
                \includegraphics[width=\textwidth]{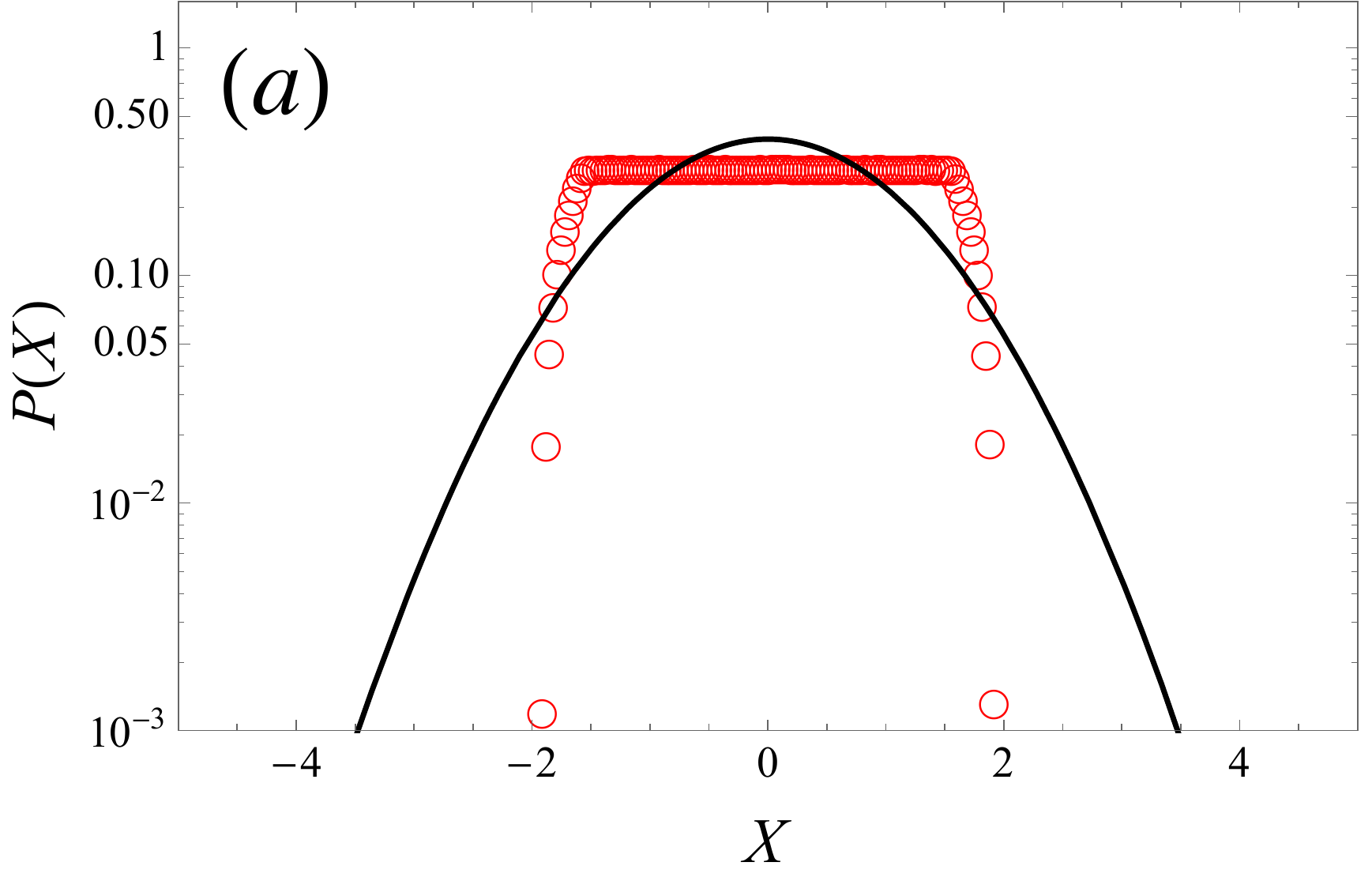}
              
        \end{subfigure} ~ 
\begin{subfigure}[b]{0.43255\textwidth}
                \includegraphics[width=\textwidth]{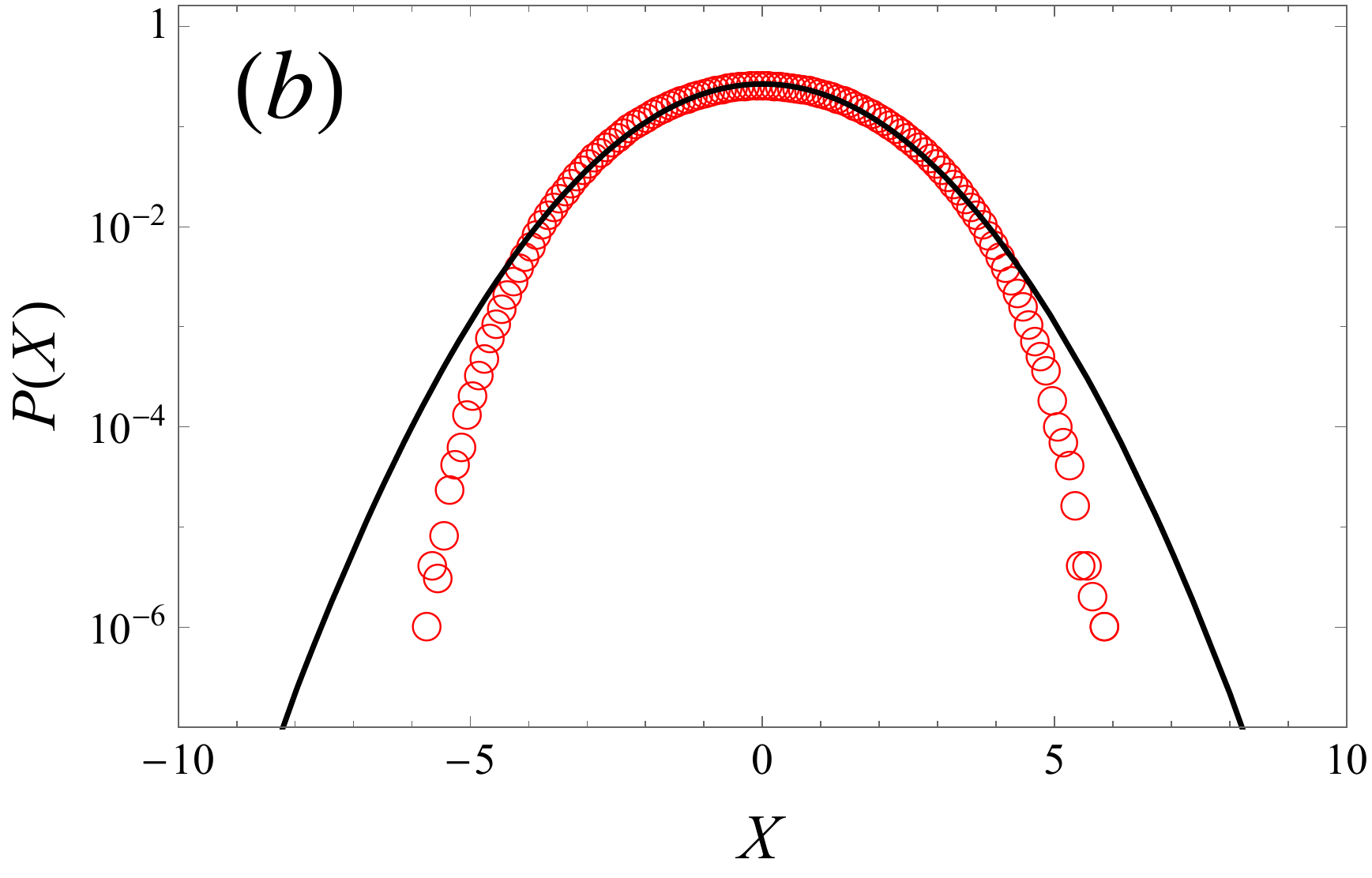}
              
        \end{subfigure} \\[0pt]
\begin{subfigure}[b]{0.4255\textwidth}
                \includegraphics[width=\textwidth]{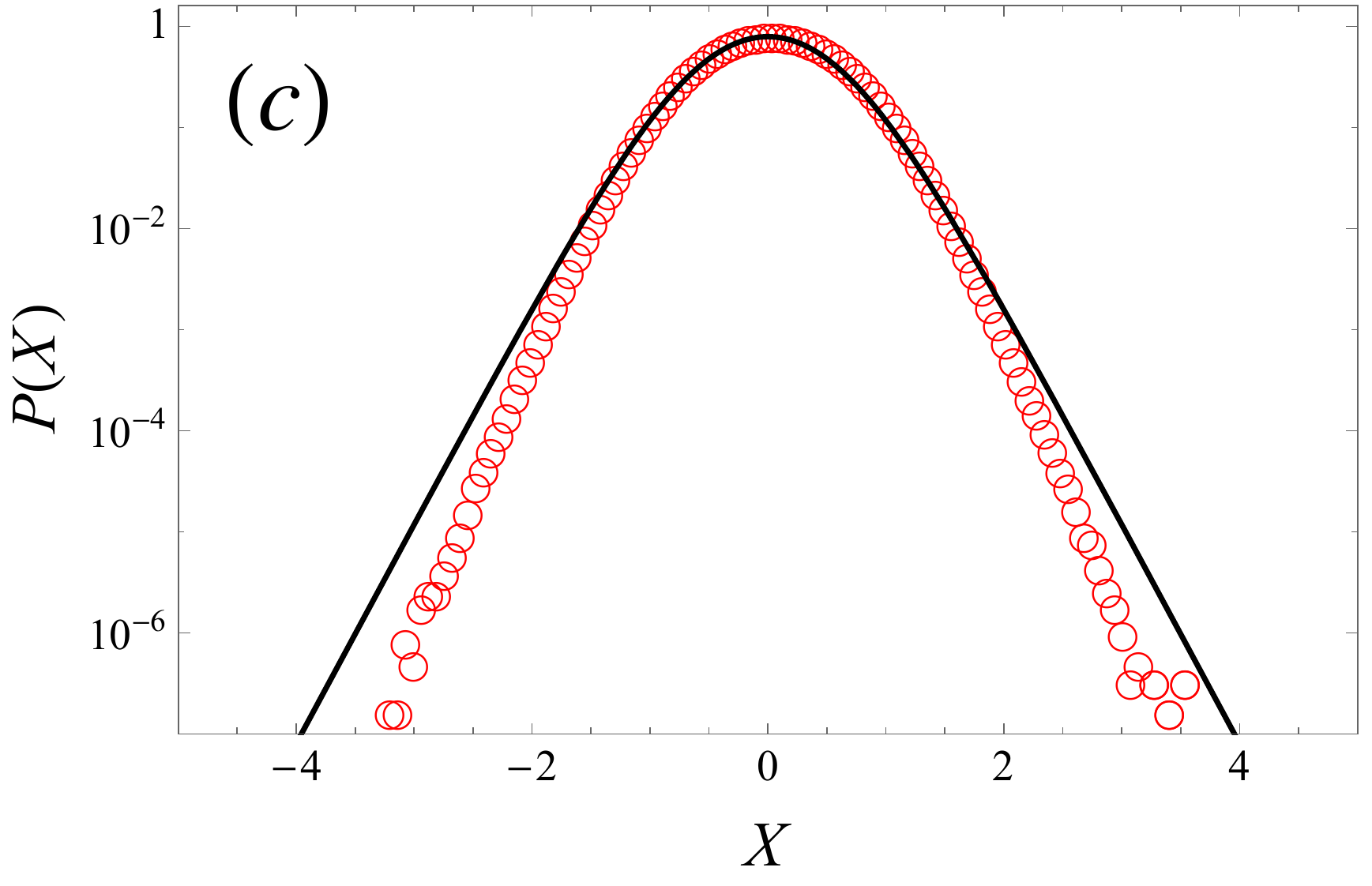}
              
        \end{subfigure} ~ 
\begin{subfigure}[b]{0.4255\textwidth}
                \includegraphics[width=\textwidth]{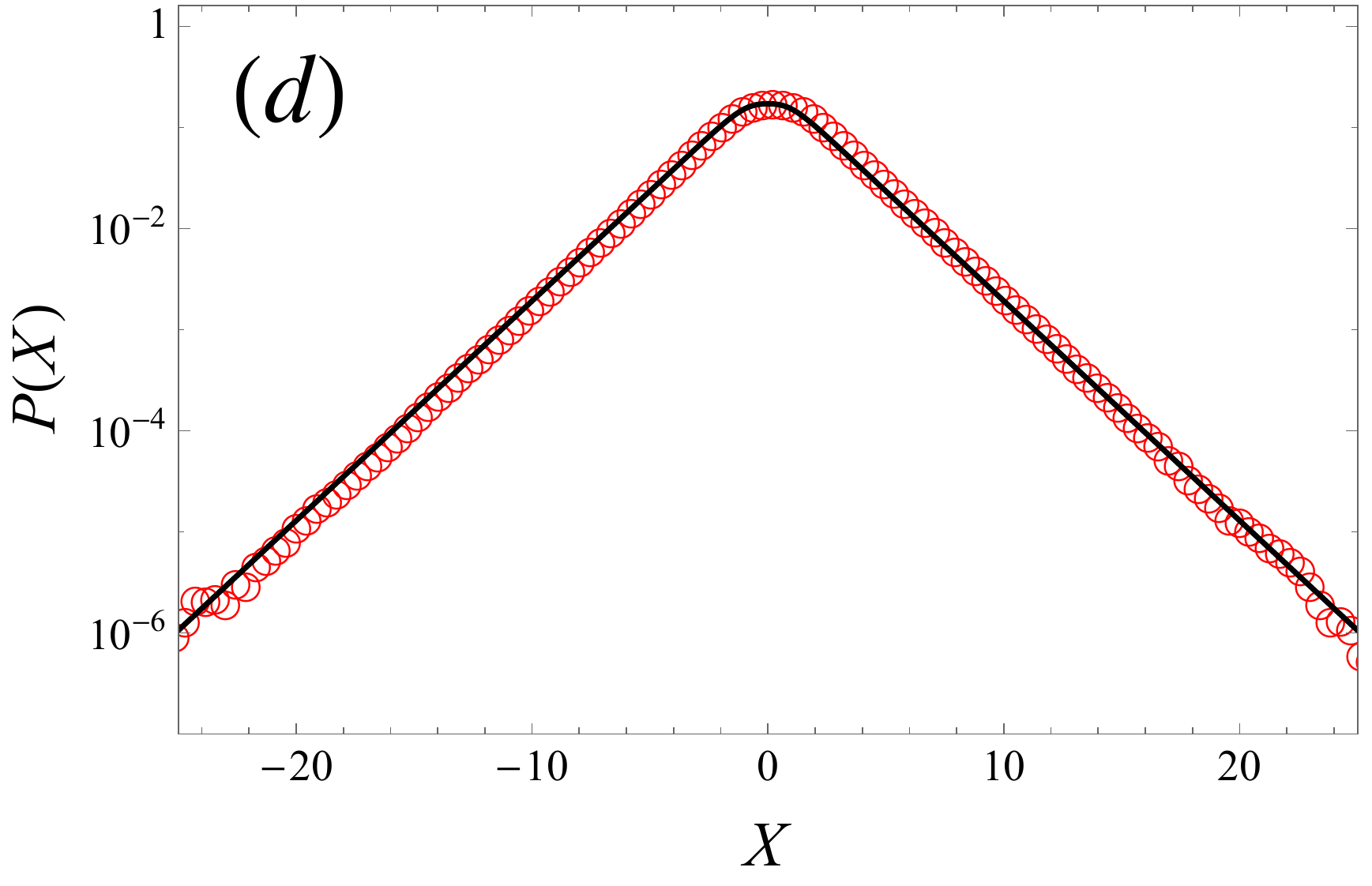}
              
        \end{subfigure}
\end{center}
\caption{ Stable distributions of stochastic maps with uniform noise, i.e. $\eta_t$ is is a random number between $-1$ and $1$ multiplied by $\beta$. The second order approximation for the stable distribution is the thick line while circles present the  simulation result. In panel {\bf (a) } the map is $a(X)=-1.1X$ and $\beta=\sqrt{3}$. In panel {\bf (b)} the map is $a(X)=-0.25X$ and $\beta=\sqrt{3}$. In panel {\bf (c)} the map is $a(X)=-0.25\tanh(X)$ and $\beta=\sqrt{3/10}$ and panel {\bf(d)} is similar to {\bf(c)} except $\beta=\sqrt{3}$.
}
\label{figflat}
\end{figure}

\begin{figure}[t]
\begin{center}
\begin{subfigure}[b]{0.4255\textwidth}
                \includegraphics[width=\textwidth]{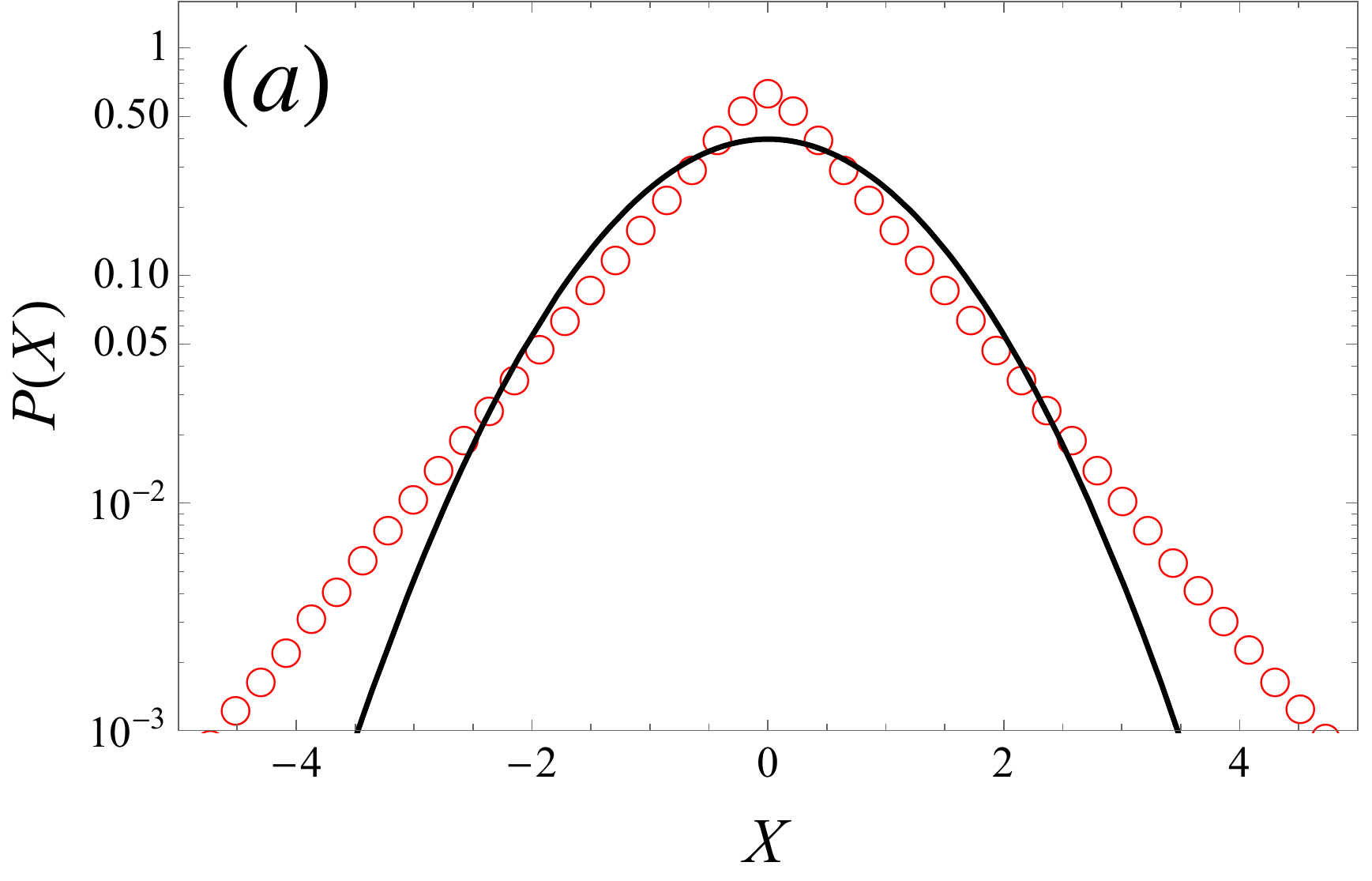}
              
        \end{subfigure} ~ 
\begin{subfigure}[b]{0.43255\textwidth}
                \includegraphics[width=\textwidth]{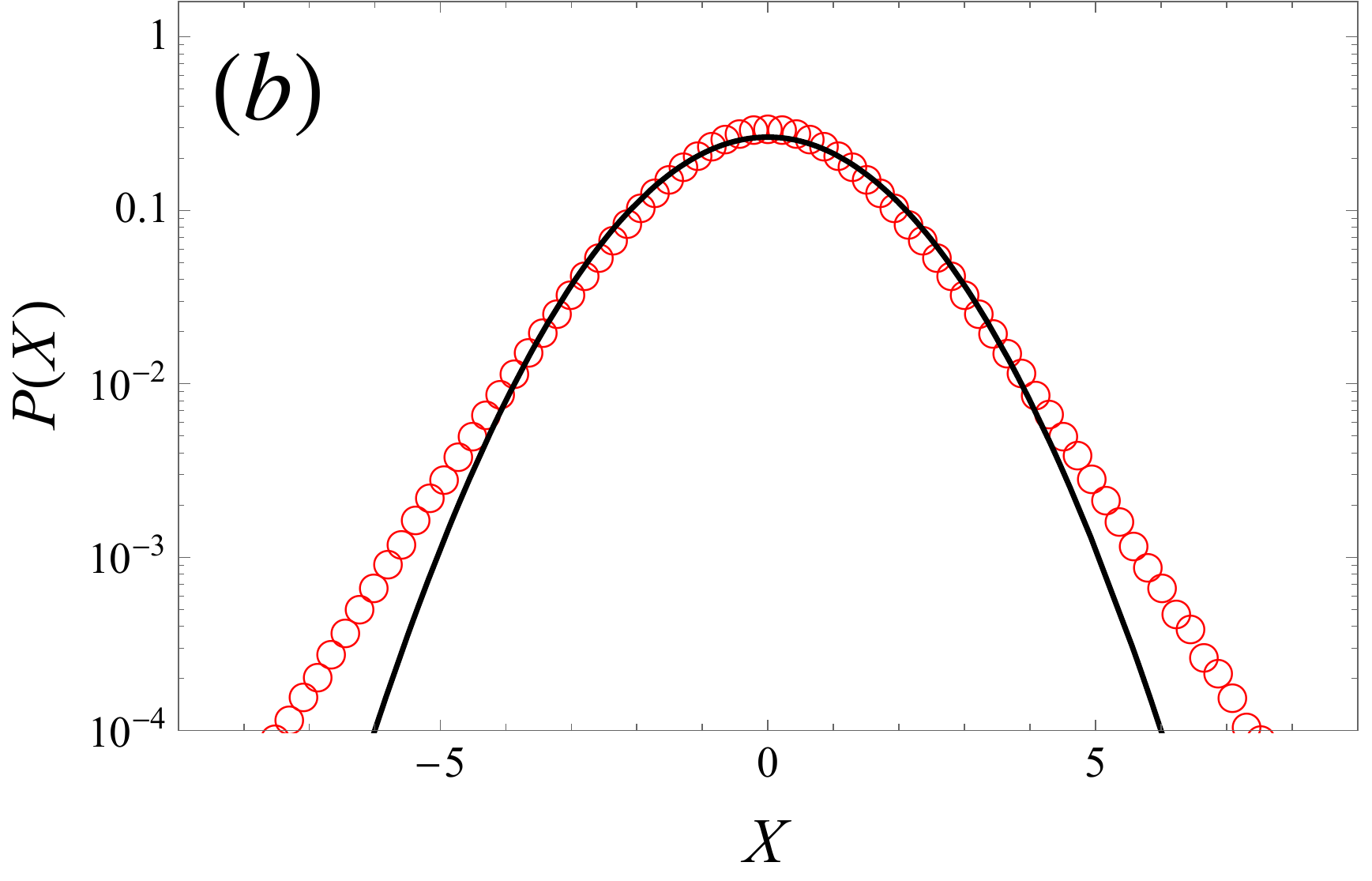}
              
        \end{subfigure} \\[0pt]
\begin{subfigure}[b]{0.4255\textwidth}
                \includegraphics[width=\textwidth]{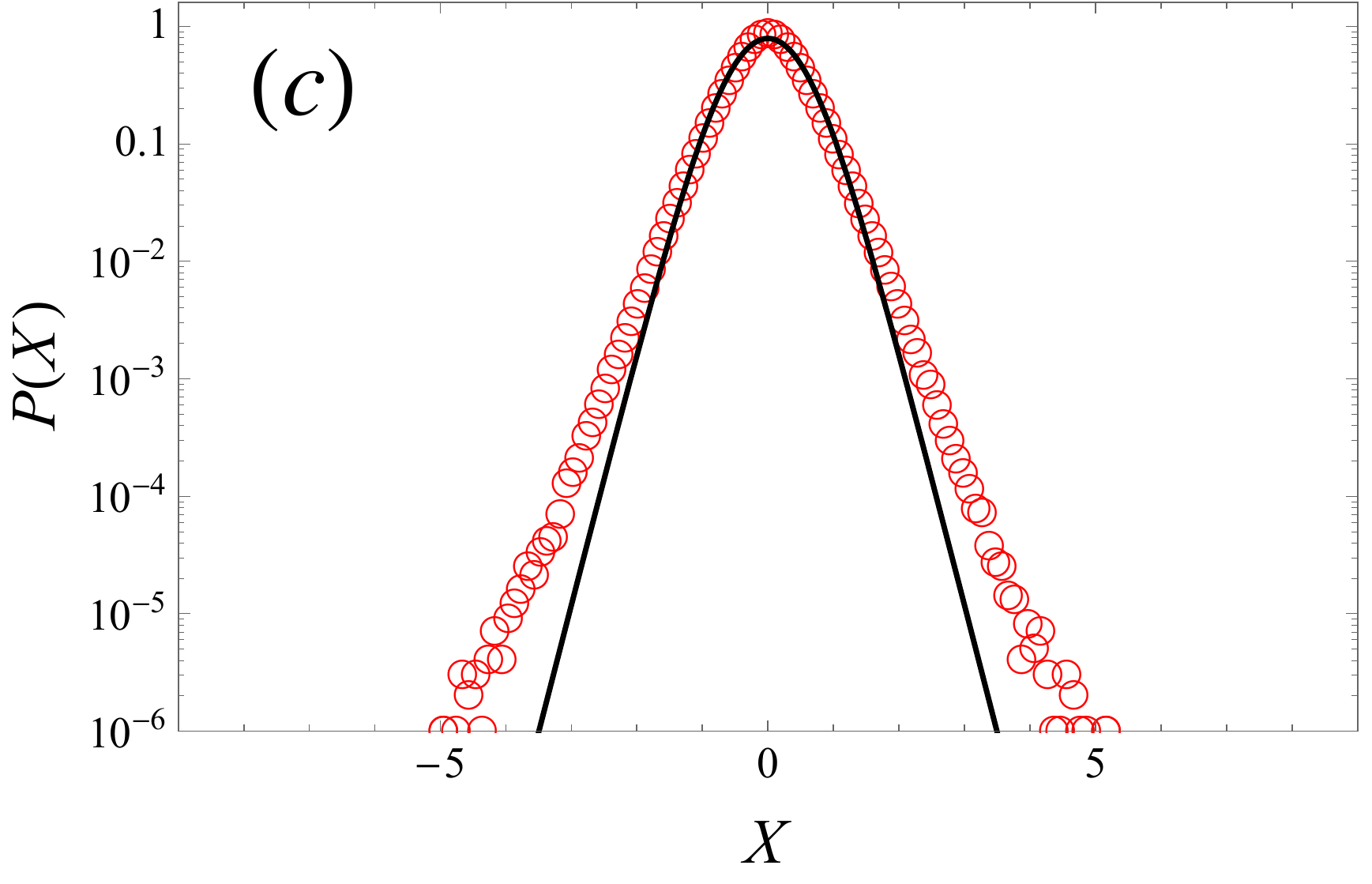}
              
        \end{subfigure} ~ 
\begin{subfigure}[b]{0.4255\textwidth}
                \includegraphics[width=\textwidth]{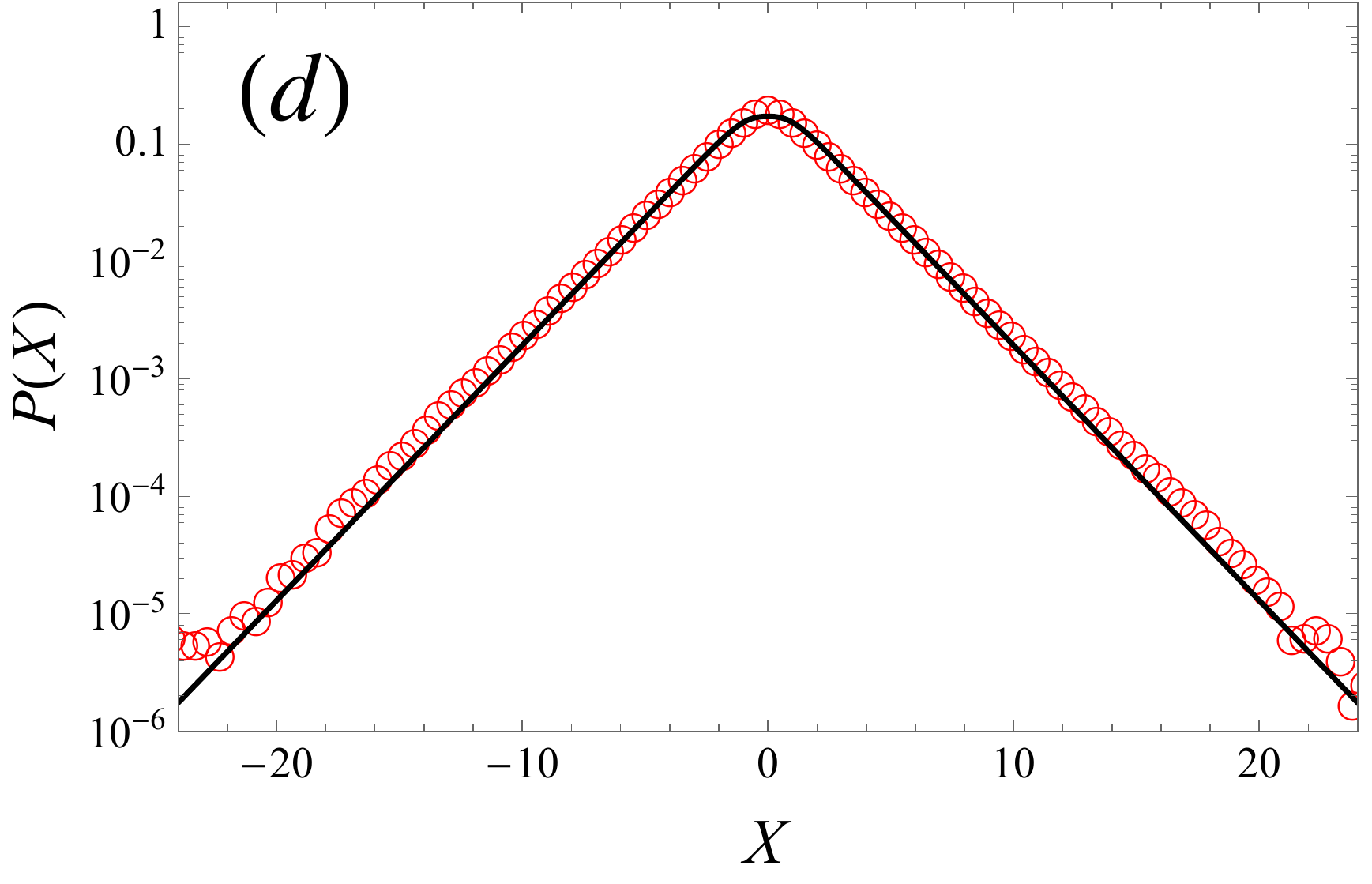}
              
        \end{subfigure}
\end{center}
\caption{ Stable distributions of stochastic maps with noise following a zero mean Laplace distribution, i.e. the probability density of $\eta_t$ is $P(\eta_t)=\frac{1}{2 \zeta}\exp\left(-|\eta_t|/\zeta\right)$ with second moment $\langle \eta^2 \rangle =2\zeta^2$. The second order approximation for the stable distribution is the thick line while circles present the  simulation result. In panel {\bf (a) } the map is $a(X)=-1.1X$ and $\langle \eta^2 \rangle=1$. In panel {\bf (b)} the map is $a(X)=-0.25X$ and $\langle \eta^2 \rangle=1/10$. In panel {\bf (c)} the map is $a(X)=-0.25\tanh(X)$ and $\langle \eta^2 \rangle=1$ and panel {\bf(d)} is similar to {\bf(c)}  except  $\langle \eta^2 \rangle=1$.
}
\label{figLap}
\end{figure}

\subsubsection{Approximation Limits}
\label{approxim}

For the SMs presented so far the continuous approximation worked well, and only for some parameters of the hyperbolic tangent map, Eq.~\ref{hyp01}, some discrepancies were observed. The discrepancy between the SM stable distribution and  $P(X)={\cal N}^{-1} \exp\left(-H(X)\right)$ with $H(X)$ provided by Eq.~(\ref{stable04}) is not coincidental. Since several assumption were invoked while deriving Eq.~(\ref{SM10}) those assumptions must be satisfied when approximating a given SM. The noise must be uncorrelated Gaussian noise and the derivative $|\partial a(X)/ \partial X|$ sufficiently small. The second condition is due to the fact that essentially the derived approximation is a series expansion truncated after two terms.  
If the local derivative is sufficiently large, both deterministic and stochastic parts of Eq.~(\ref{SM10}) are unbounded. At the level of the stable distribution, the fact that we used a series expansion suggests that the correction term, i.e. the $\ln$ term on the r.h.s of Eq.~(\ref{stable04}), should be small. When the map derivative is large so is the $\ln$ term and noticable discrepancies between the true and approximated behavior are expected. 
Explicitly, for the hyperbolic tangent map the the presented discrepancy in Fig.~\ref{fighyp} appeared when $z=1.5$ and the local derivative became larger then $1$. The fast local changes, rather then the global non-linear features, are responsible for the loss of precision of the approximation. For an oscillating map such as 
$a(X)=-0.25\tanh(X)+0.5\cos(X)$ and noise $\langle \eta_t ^2\rangle =0.4^2$ the approximation works extremely well, see Fig.~\ref{figoscill} panel (a). When the map can change locally quite quickly the approximation is not as good anymore. For the oscillating map example, when the $\cos$ part was changed to $0.5\cos(2X)$ the approximation began to break down, see Fig.~\ref{figoscill} panel (b). 
This limitation of sufficiently small changes in $a(X)$ is reasonable from the perspective of the derivative series approximation. Keeping more terms of the series is expected to improve the situation.

The second limitation is non-correlated Gaussian noise. The approximation was developed under this strict assumption which allows the derivation of Eq.~(\ref{SMR})  and other kind of noises can't be simply represented as a Wiener integral. The presented approximation has no way to account for the non-Gaussian nature of the noise. Under some circumstances one can expect that Gaussian noise approximation to be sufficient, since a sum of random variables is well approximated by a Gaussian distribution, given that the first and second moments exist. However, during iterative application of the SM the noise $\eta_t$ is not simply  a sum of uncorrelated random variables, and so this argument does not hold. In Figures \ref{figflat} and \ref{figLap} we present the SM behavior when the noise distribution is uniform in the domain $(-1,1)$  and given by a Laplace distribution, respectively.
 The discrepancies between the true behavior and continuous approximation are obvious and depend not only on the noise strength and variance but also on the deterministic part of the map.

\begin{figure}[t]
\begin{center}
\begin{subfigure}[b]{0.47\textwidth}
                \includegraphics[width=\textwidth]{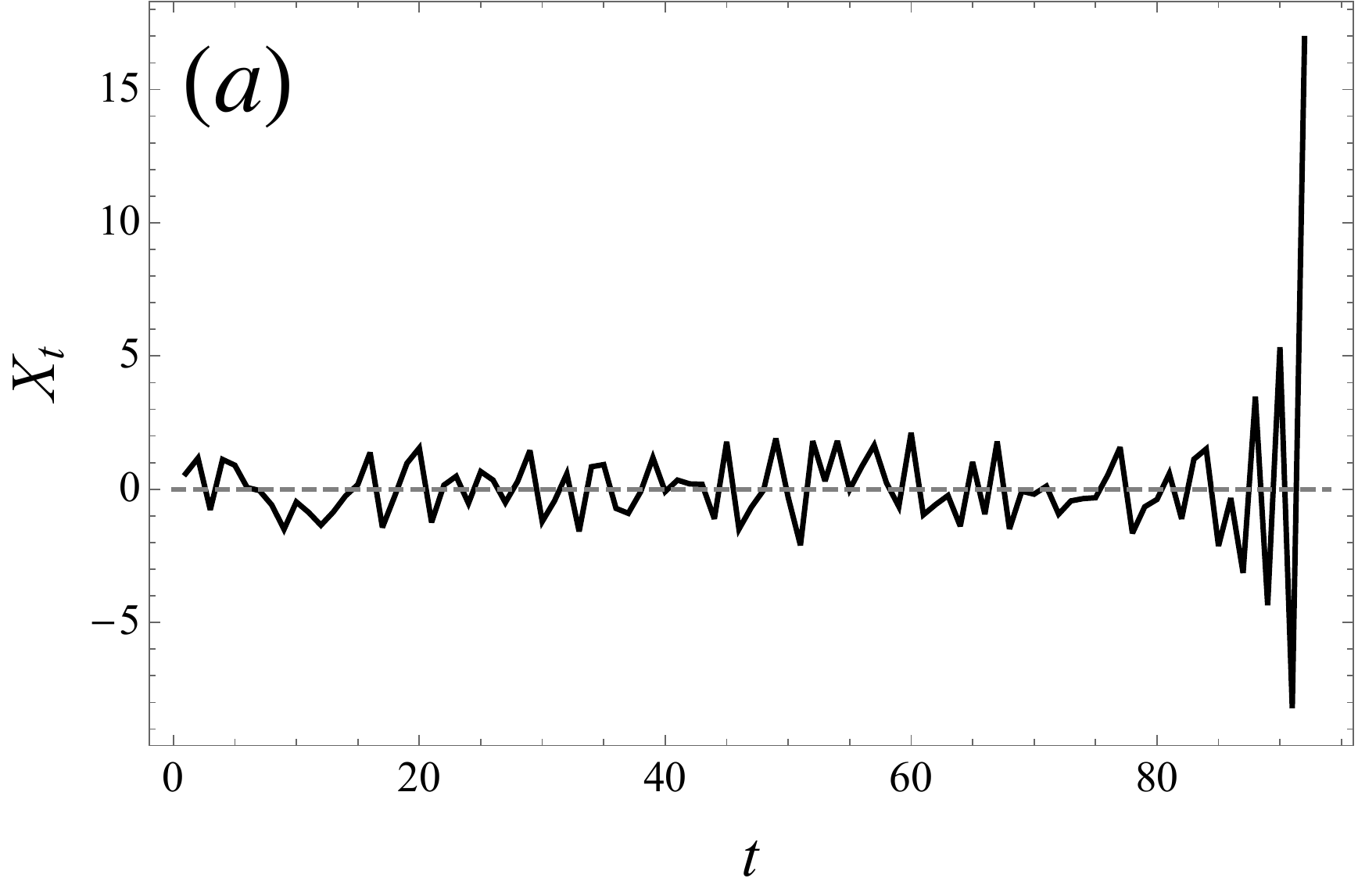}
              
        \end{subfigure} ~ 
\begin{subfigure}[b]{0.47\textwidth}
                \includegraphics[width=\textwidth]{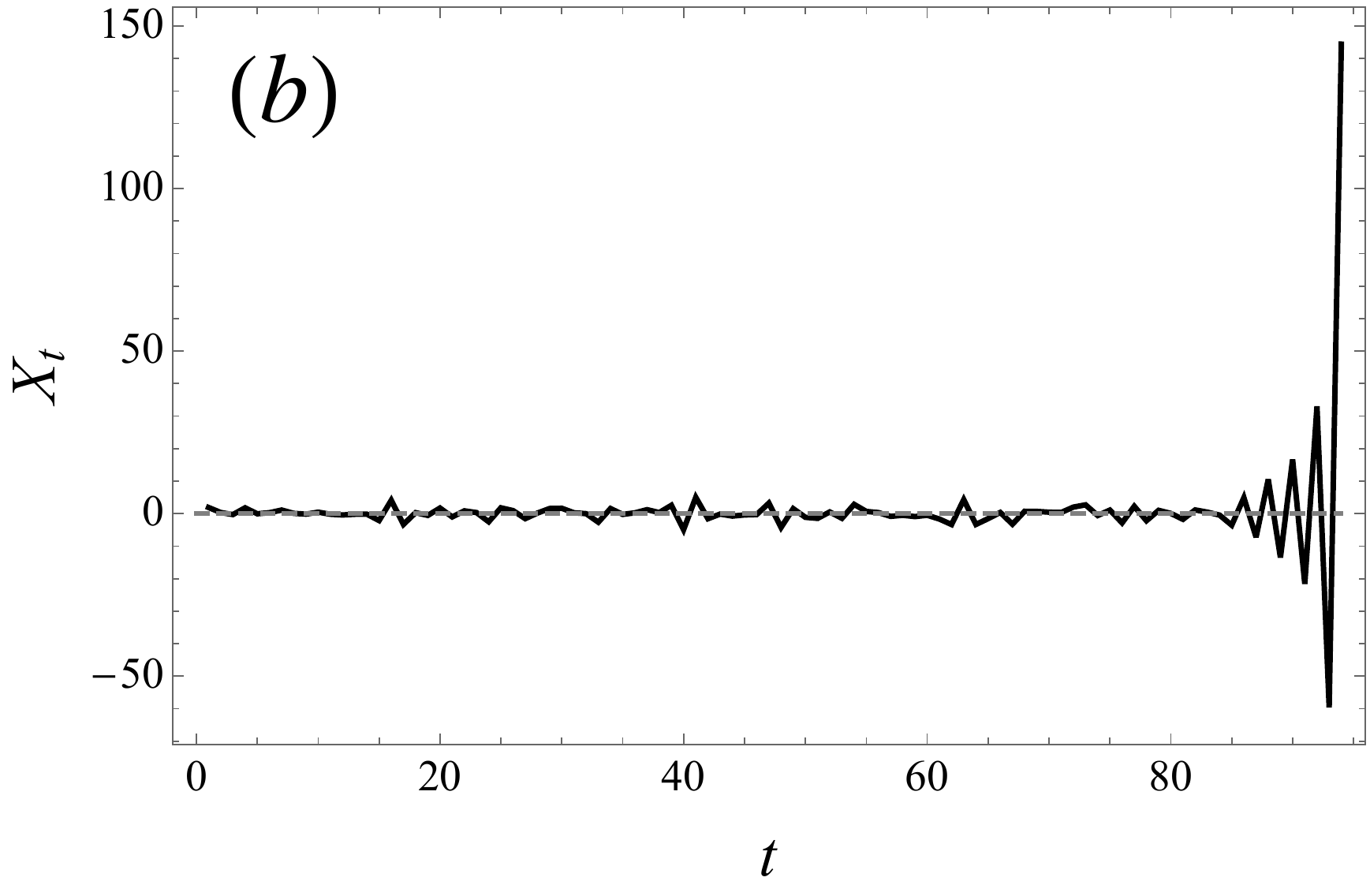}
              
        \end{subfigure} \\[0pt]
%
%
\end{center}
\caption{ Temporal traces of unstable maps with Gaussian noise. Panel {\bf (a)} present a single trace for $a(X)=-X|X|^{1/2}$ and $\langle \eta^2 \rangle=1$. Panel  {\bf (b)} present a single trace for $a(X)=-X|X|^{3/10}$ and $\langle \eta^2 \rangle=3$. For both cases the trajectory escapes to infinity in an oscillating fashion.
}
\label{figunstable}
\end{figure}

\subsection{Stability criterion}
\label{stabilityC}

In the previous subsection, we dealt with the stable distribution of the SM and our approximation to it.  The condition for an SM to attain a stable distribution is not obvious when considering the map itself. The continuous approximation, on the other hand, provides a simple prescription for the existence of a stable distribution. From Eq.~(\ref{stable02}) and Eq.~(\ref{stable04}) it is clear that the existence of a stable solution depends on the form of $H(X)$. Since the solution is of the form $ \exp\left(-H(X)\right)$, $H(X)$ must stay greater than $0$ for $X\to\pm\infty$.  Otherwise, the solution is non-normalizable, i.e., ${\cal N}\to\infty$, and no stationary solution exists. 
The form of $H(X)$ is provided by Eq.~(\ref{stable04}) and in order to fulfill this condition the integral of $a(X)$ must be negative and satisfy $\int a(X)\,dX<-\frac{1}{4}a(X)^2$ for sufficiently large $|X|$. This sets an upper bound for the behavior of $|a(X)|$ as $X\to\pm\infty$. The limit growth of $|a(x)|$ is $|2X|$. For any function that is growing faster than (or at the same rate as) $|2X|$ we expect to obtain  unstable behavior. 
In Fig.~\ref{figunstable} two cases of unstable behavior are presented. A stable distribution is unachievable since any given trajectory $X_t$ will eventually escape to infinity. The time of escape will vary but, eventually, a scenario similar to the one presented in Fig.~\ref{figunstable} will develop.

The ``physical" reasoning behind the proposed limit is as follows. The SDE equation provided by Eq.~(\ref{SM10}) can be viewed as a Langevin Equation displaying the behavior of a particle in an effective potential provided by $H(X)$.  
The question of stability then boils down to the binding properties of the potential. When $H(X)$ is non-binding  the effective dynamics is governed by trajectories that escape to infinity.  It may happen that locally $H(X)$ has a minimum and if the noise is not significantly high the process will stay for a very long time in the vicinity of this minimum. Still, due to fluctuations, the particle is bound to escape.
In the stability criterion we have derived, only the first two terms of $H(X)$ were exploited. It has been already shown that the correction term, i.e. the $\ln(|\dots|)$, can take large values and be the cause of large discrepancies between the approximation and the true behavior (see Sec.~\ref{approxim}). From the  examples presented, we see that the discrepancies are local (for Gaussian noise) while the stability criterion is based on the asymptotic behavior of $H(X)$. The series truncation that was performed while deriving Eq.~(\ref{SM10}) significantly changes $H(X)$ and thus can affect  its local shape and thereby determine the time-scale of escape to infinity (``physically" speaking, the height and existence of local potential barriers). However, since all the corrections are functions of derivatives of $a(X)$ we conjecture that only    
$ -2\int a(X)\,dX-a(X)^2 /2$ determines the asymptotic binding properties of $H(X)$.

\section{Summary}
We have presented a systematic approach for the continuous approximation of discrete maps, both stochastic and non-stochastic. For the non-stochastic case, we obtained an approximation which describes the temporal evolution of the map. The comparison to a linear map and the Pomeau-Manneville map shows good agreement between our approximation and the true behavior (obtained numerically). For the stochastic case we utilized the It\^o stochastic calculus in order to approximate the discrete map as a type of Langevin equation. In the case where the map is non-linear, the second order approximation yields an equation with the presence of multiplicative noise. We derived an equation for the stable distribution of an SM with uncorrelated Gaussian noise and compared it to numerically obtained distributions of several SMs . The multiplicative  noise can  manifest itself as an effective coupling between the deterministic part of the map and the noise strength. It might be physically very important to understand when the process ceases to describe a stable situation. The derived continuous approximation provides a simple criterion for stability of SM depending on  the normalizability of the obtained distribution.

While the approximation works quite well for moderate non-linearity and noise values, due to the perturbative nature of the approximation it can break down imn more extreme cases. Indeed we have seen that the second order SM stops working when the local jumps (of the map evolution) become sufficiently large.  Further exploration of the presented approximation carried out to higher orders is needed.
We expect that the presented results will become quite valuable in any field where noise and discreteness of evolution parameter is essential, for example, cell division, since  cell divisions are discrete events~\cite{Sri,SookJoon,Soifer,Tanuchi,Balaban,HSalman}.

\bibliographystyle{ieeetr}
\bibliography{maplit}

\end{document}